\documentclass{article}

\usepackage[english]{babel}

\usepackage{amsfonts}               
\usepackage{amsmath}	            
\usepackage{mathtools}              
\usepackage{amssymb}                
\usepackage{amsthm}                 
\usepackage[shortlabels]{enumitem}	
\usepackage{booktabs}

\usepackage{bbold}                  

\usepackage{tcolorbox}
\newtcolorbox{thm}{center,
    colframe=white,
    colback=black!10,
    left=1pt,
    right=1pt,
    top=1pt,
    bottom=1pt
}

\newtheorem{definition}{Definition}
\newtheorem{proposition}[definition]{Proposition}
\newtheorem{remark}[definition]{Remark}
\newtheorem{corollary}[definition]{Corollary}

\newcommand{\pol}{\mathsf{\pi}}     
\newcommand{\dec}{\tau}       
\newcommand{\x}{\mathsf{x}}       
\newcommand{\y}{\mathsf{y}}       
\newcommand{\prop}{e}            
\newcommand{\proptU}{\prop_{t,\Xpart}}  

\newcommand{\ytil}{\Tilde{\y}}       
\newcommand{\yitil}{\Tilde{y}_i}       

\newcommand{\eps}{\epsilon}         

\newcommand{\py}{p}            
\newcommand{\s}{\mathsf{t}}            
\newcommand{\Xpart}{\Pi}            

\newcommand{\qtpHat}{\hat{\xi}^t_p}
\newcommand{\qtp}{\xi^t_p}
\newcommand{\qtpJJ}{\xi^{\JJ_t}_p}

\newcommand{\avgIIpt}{\rho(\II, t)}
\newcommand{\avgJJpt}{\rho(\JJ, t)}
\newcommand{\avgIIyt}{\mu(\II, t)}

\newcommand{\avgIIytx}{\mu_x(\II, t)}
\newcommand{\avgJJytx}{\mu_x(\JJ, t)}
\newcommand{\avgIIytU}{\mu_U(\II, t)}
\newcommand{\avgJJytU}{\mu_U(\JJ, t)}

\newcommand{\ZZ}{\mathcal{Z}}
\newcommand{\XX}{\mathcal{X}}
\newcommand{\YY}{\mathcal{Y}}

\newcommand{\GG}{\mathcal{G}}
\newcommand{\II}{\mathcal{I}}
\newcommand{\JJ}{\mathcal{J}}

\newcommand{\Scal}{\mathcal{S}}
\newcommand{\TT}{\mathcal{T}}

\newcommand{\RR}{\mathbb{R}}

\newcommand{\EE}{\mathbb{E}}

\newcommand{\indep}{\perp\!\!\!\perp}

\newcommand{\littletaller}{\mathchoice{\vphantom{\big|}}{}{}{}}
\newcommand\restr[2]{{
  \left.\kern-\nulldelimiterspace 
  #1 
  \littletaller 
  \right|_{#2} 
  }}

\usepackage[bibstyle=authoryear, citestyle=authoryear, backend=biber, natbib]{biblatex}
\bibliography{library}

\title{Formalising causal inference as prediction on a target population}

\author{Benedikt Höltgen \quad\quad Robert C. Williamson}
\date{\normalsize University of Tübingen and Tübingen AI Center}

\begin{document}
\maketitle

\begin{abstract}
    The standard approach to causal modelling especially in social and health sciences is the potential outcomes framework due to Neyman and Rubin. In this framework, observations are thought to be drawn from a distribution over variables of interest, and the goal is to identify parameters of this distribution. Even though the stated goal is often to inform decision making on some target population, there is no straightforward way to include these target populations in the framework. Instead of modelling the relationship between the observed sample and the target population, the inductive assumptions in this framework take the form of abstract sampling and independence assumptions. 
    In this paper, we develop a version of this framework that construes causal inference as treatment-wise predictions for finite populations where all assumptions are testable in retrospect; this means that one can not only test predictions themselves (without any fundamental problem) but also investigate sources of error when they fail. Due to close connections to the original framework, established methods can still be be analysed under the new framework.
\end{abstract}


\section{Introduction}

For many problems in the social and health sciences, it is important to analyse and predict the efficacy of treatments or policies.
This requires causal modelling, as observational outcome distributions may not reflect outcome distributions under active treatment policies.
The Potential Outcome framework due to Neyman \citep{neyman1923} and \citet{rubin1974} is the dominant approach to causal modelling in many fields.
Despite Neyman's early work, we shall follow \citep{holland1986} in calling these models Rubin causal models (RCMs), as we often specifically refer to Rubin's now-dominant version that is framed in terms of probability distributions.
RCMs are the preferred framework in particular for informing specific policies or interventions, due to the focus on specific outcomes of interest and the aptitude to accommodate individual problem settings \citep{imbens2020, markus2021} -- \emph{vis-a-vis} the standard econometric approach \citep{heckman2024} and structural equation models \citep{pearl2009}, which are are sometimes seen as having the edge in more abstract theory building, including the modelling of unobservables.
However, while the `credibility revolution’ of the last decades resulted in theoretically rigorous methods \citep{angrist2010}, it `has focused primarily on internal validity’ \citep[p. 1070]{egami2023}.
In contrast, while already the original paper introducing RCMs called on `investigators [to] carefully describe their sample of trials and the ways in which they may differ from those in the target population' \citep[p. 698]{rubin1974}, the analysis of external validity has been widely neglected (as attested by lamentations across fields, see Section~\ref{ss:target}).

We argue that this has in part to do with the framework itself, as it does not provide a straightforward way to model target populations.
Instead of effects on target populations, the focus of the framework is the `identification' of parameters in some abstract probability distribution -- `as if a parameter, once well established, can be expected to be invariant across settings’ \citep[p. 10]{deaton2018}.
A further issue with this abstract framing is that it is very difficult, if possible at all, to formulate concrete and testable assumptions that enable the accurate prediction of the effects of policies on target populations.
In this paper, we suggest an amendment to the framework to overcome, or at least mitigate, these problems.
More concretely, we suggest a variant of RCMs that directly models both observed and target populations and their relationship, avoiding detours through abstract distributions.
It shifts the focus from the identification of true parameters to the prediction of future outcomes based on (retrospectively) testable assumptions.
Rather than ignoring existing causal inference methodology, we show that the new framework can capture established estimators and provide a complementary perspective on them.
We, thus, provide an `intermediary' framework that establishes links between high-level intuitions, as formalised in RCMs, on the one hand and directly testable assumptions about concrete populations on the other.
Hence, this variant augments the strengths of RCMs for evidence-based policy making by focussing on predictions for concrete target populations grounded in testable assumptions.
Beyond these benefits, it also offers complementary perspectives not only on established causal estimators but also on causal inference as a whole.

The structure of this work is as follows.
In Section~\ref{s:RCMs}, we recapitulate RCMs and discuss calls across fields to direct more attention to problems with external validity and to model the target population more directly.
In Section~\ref{s:new_frame}, we introduce the new framework in the context of simple estimators; a broader survey of existing estimators and how they fit into the new framework can be found in Appendix~\ref{app:new_persp}.
In Section~\ref{s:compare}, we compare the new with the conventional framework, by drawing formal connections and discussing differences in practice.
While the bulk of the paper focuses on \emph{average} treatment effects and potential outcomes, Section~\ref{s:beyond_apo}, briefly addresses generalisations to conditional treatment rules as well as to distributional properties beyond the mean.
Section~\ref{s:discussion} concludes and discusses how the framework how the framework accommodates a less metaphysically loaded view on causal inference as a whole.


\section{Background: Rubin Causal Models (RCMs)}
\label{s:RCMs}

In this section, we first introduce the standard formalism of RCMs, before critically discussing the assumption of an underlying probability distribution as well as the problem of explicitly modelling outcomes on a target population.


\subsection{The framework}

The main components of RCMs are the following random variables (RVs):
$T_i$ is the decision variable indicating whether person $i$ is treated and takes values in $\{0,1\}$, which represent control and treatment.\footnote{We only consider the binary treatment case here as this makes the presentation of RCMs simpler.}
$Y_{1i}$ and $Y_{0i}$ denote the outcome for $i$ upon receiving treatment and control, respectively.
Based on this, we can define the actual outcome
\begin{equation}
    Y_i := T_i \cdot Y_{1i} + (1-T_i) \cdot Y_{0i} = Y_{0i} + T_i (Y_{1i} - Y_{0i}).
\end{equation}
%
In the example of job trainings, $T_i$ indicates whether someone gets offered job training and $Y_i$ indicates whether they have a job after a fixed time, say, one year.
It is, thus, assumed that for every person, both $Y_{0i}$ and $Y_{1i}$ are well-defined, i.e., whether they find a job \textit{if} they don't get offered job training and whether they find a job \textit{if} they are assigned job training, respectively.
Of course, we can, in principle, only observe the value of one of the two variables for each individual.

In many settings (some of which we will consider in this paper), we also have informative covariates $X_i$.
In the context of job trainings, these may be attributes such as age, gender, education, and employment history.
A foundational assumption behind RCMs is that there is a joint distribution $P$ over all variables, i.e.
\begin{equation}\label{eq:joint_distr}
    Y_{1i}, Y_{0i}, T_i, X_i \sim P.
\end{equation}
We are then usually interested in the average treatment effect (ATE) $\EE_P[Y_{1i} - Y_{0i}]$.
The ATE is typically seen as the expectation over the individual treatment effect (ITE) $Y_{1i} - Y_{0i}$.
The fact that we can only ever measure one of them for each $i$ has been dubbed the `fundamental problem of causal inference'.

In line with many textbooks, we showcase RCMs in the context of Randomised Controlled Trials (RCTs).
This represents the `experimental ideal' in the sense that it assumes we have access to data from a randomised experiment.
We can then assume that the potential outcomes are independent of the treatment decision
\begin{equation}\label{eq:random_ass}
    Y_{1i}, Y_{0i} \indep T_i.
\end{equation}
This means we don't need covariates $X_i$ here, as the ATE can be expressed as
\begin{equation}\label{eq:E-E}
    \EE_P[Y_{1i} - Y_{0i}] = \EE_P[Y_i | T_i=1] - \EE_P[Y_i | T_i=0],
\end{equation}
and both terms on the RHS can directly be estimated from our data:
Assuming that past and future data are sampled from the distribution $P$, the law of large numbers says that the empirical estimate of (\ref{eq:E-E}) converges to the ATE almost surely.

RCTs is are often not available and many methods have been devised to allow causal inference when random assigment and thus the conditional independence (\ref{eq:random_ass}) is violated.
Most of these methods rely on another assumption called \textit{unconfoundedness} (also conditional independence, ignorability, selection-on-observables), given by
\begin{equation}\label{eq:unconf}
    Y_{1i}, Y_{0i} \indep T_i | X_i.
\end{equation}
This means that treatment assignment may have been based on $X_i$ but not on any other variables that could give information about the outcome.
In other words, the treatment group should be comparable to the control group if we take the covariates into account.
For example, in labour market programmes, information about potential participants is taken into account for deciding who gets offered job training.
The unconfoundedness assumption is is then assumed to be satisfied if the decision was only based on the covariates $X_i$ -- hence the name `selection on observables'.

\subsection{The distribution}
\label{ss:distribution}

As described above, RCMs are formulated in terms of joint distributions that represent the ground truth and are used to derive theoretical guarantees.
But how should these distributions be understood?
Should we take them at face value and believe that they are supposed to correctly describe some data-generating process?
Or are they just models that have a more indirect relationship with what we believe to be true about the world?
We now discuss these two options in turn.

The first option is that descriptions in terms of generative distributions can be true or false.
Much of the literature suggests this reading.
This would mean e.g. that there is a true (marginal) distribution of covariates from which people are sampled.
And this seems to hold then for any selection of covariates/attributes that we can come up with.
In principle, the number of possible descriptions (choices of covariates) with is almost unlimited, although we often just take the attributes that we can most easily measure.
The invoked distributions are commonly not thought to pertain to a specific time, though sometimes to a specific location.
But the relationships between e.g. unemployment, age, and education clearly change over time -- if they can be said to be stable at any given point in the first place.
Every case of a person finding or not finding a job is highly individual and it seems rather strong that they just follow a general law plus some noise -- a notion we know from physics.
Nancy \citet{cartwright1999} makes the point that while a lot of work in physics has focused on finding latent quantities that do have stable relationships (like forces in Nowton's and Hooke's laws), social sciences like economics typically consider more easily measurable quantities that are particularly salient.\footnote{This applies particularly to RCMs which, in contrast to other causal modelling approaches \citep{heckman2024} do not model unobservables.}
And `to suppose that there really is some probability measure over [such quantities], you need a lot of good arguments’  (p. 325).

The other option is, then, to say that aspects like sampling from a joint distributions are mere models that are always wrong (if they have truth conditions at all).
On this view, assumed joint distributions are idealisations and aim to capture patterns that we can observe between individual events, but they cannot be literally true.
Arguments of this kind go back at least to de Finetti, who showed that Bayesians with certain priors can equivalently describe their subjective credences as if they were sampling i.i.d. from imaginary distributions.
If this is to be the interpretation, it is surprising how little attention has been paid to how the idealised model relates to the world we observe.
What, for example, does the assumption of unconfoundedness mean if it can never be true?
If generative distributions are useful fictions, under which conditions are they useful?
While a realist about distributions should already provide bridges to observable statements of interest (beyond the invocation of i.i.d. samples), an instrumentalist has all the more reason to do so.

\begin{figure}[ht]
    \centering
    \includegraphics[width=0.95\linewidth]{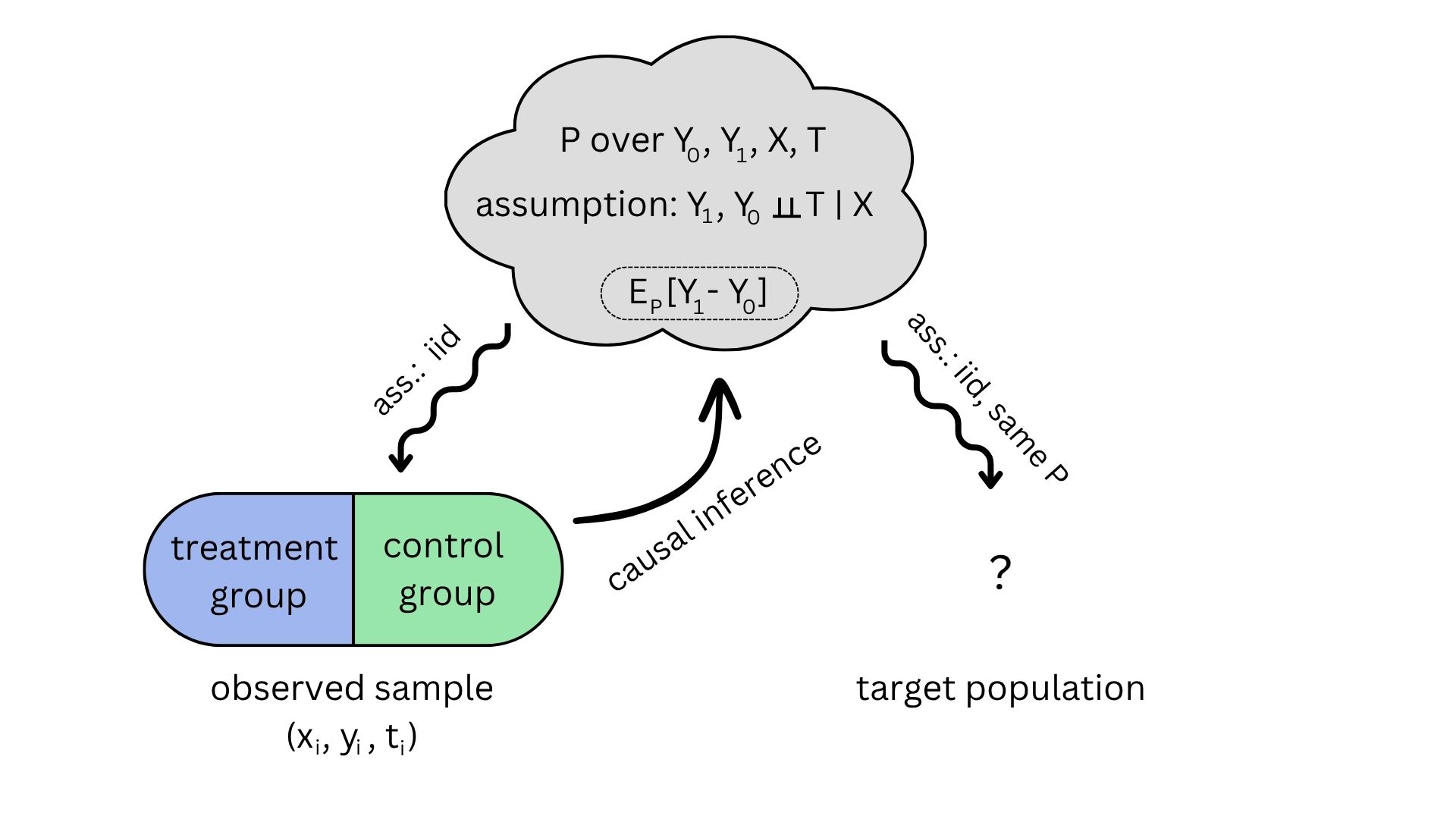}
    \caption{A schematic drawing of the RCM approach.}
    \label{fig:RCMs}
\end{figure}

While a standard and seemingly innocuous aspect, these distributions actually do a lot of the heavy lifting.
They provide the connections between both the quantities of interest and between different populations, via sampling assumptions.
Getting the former right requires internal validity while getting the latter right requires external validity; in RCMs, these two aims are typically considered in separation.
Internal validity is concerned with the assumptions discussed above, especially with whether unconfoundedness holds in the assumed `generative process' behind observational data.
This is supposed to ensure that estimations of quantities like the ATE are correct for the observed sample.
External validity, in contrast, concerns the question whether these estimations also hold for future data, that is, for populations on which we want to make treatment decisions in the future.
It is usually assumed that the distribution reflected in our historical data is the same as or similar to the distribution from which future data is `sampled', which allows us to predict outcomes or treatment effects in some population of interest.
Figure~\ref{fig:RCMs} provides a visual sketch of the RCM picture, where the top and left parts (distribution and observed population) cover internal validity.

\subsection{The target population}
\label{ss:target}

‘For almost any study to be of interest, the results must be generalizable to a population of trials.’ This quote comes from Rubin's first paper introducing the framework \citep[p. 699]{rubin1974}.
As he also noted, 
\begin{quote}
    `in order to generalize the results of any experiment to future trials of interest, we minimally must believe that there is a similarity of effects across time and more often must believe that the trials in the study are ``representative'' of the population of trials. [...] Even though the trials in an experiment are often not very representative of the trials of interest, investigators do make and must be willing to make this assumption [...] in order to believe their results are useful.’ \citep[p. 698]{rubin1974}.
\end{quote}

However, it has been often noted that `[s]ocial scientists frequently invoke external validity as an ideal, but they rarely attempt to make rigorous, credible external validity inferences.’ \citep[p. 365]{findley2021}.
It is striking how pervasive such statements are across various fields in which RCMs are used.
In economics, `[r]esearchers tend to focus primarily on threats to internal validity' \citep[p. 274]{bo2021} whereas external validity is virtually not discussed in standard textbooks like \citet{angrist2010}.
In epidemiology, `threats to external validity are less well-understood’ as it is considered `the common view that external validity is secondary to, or contingent on, internal validity' \citep[p. 2]{lesko2020}.
In public health, 
`[t]he consequence of this emphasis on internal validity has been a lack of attention to and information about external validity, which has contributed to our failure to translate research into public health practice' \citep[p. 9]{steckler2008}.
In behavioral medicine, it has been observed that `[t]he majority of intervention studies conducted and reported in Annals [of Behavioral Medicine] and other health journals [...] are usually silent on external validity’ \citep[p. 106]{glasgow2006}.
Similarly, `Only 11\% of all experimental studies and 13\% of all observational causal studies published in the American Political Science Review from 2015 to 2019 contain a formal analysis of external validity in the main text, and none discuss conditions under which generalization is credible’ \citep[p. 1070]{egami2023}.
This lack of engagement with external validity across fields is problematic, especially given that in common settings, internal validity in a sense `carries much less information than the external validity assumptions’ \citep[p. 1359]{breskin2019}.

We believe that the RCM framework and its focus on identification strategies has contributed to this development; as noted by \citet[p. 1070]{egami2023}, `the credibility revolution [...] has focused primarily on internal validity’.
One reason for this arguably lies in the very notion of `identification' of parameters that define the underlying distribution -- `as if a parameter, once well established, can be expected to be invariant across settings’ \citep[p. 10]{deaton2018}.\footnote{Note however, that this problem is not unique to RCMs; indeed, advocacy of quick generalization `from the actual study experience to the abstract, with no referent in place or time’, \citep[p. 47]{miettinen1985} predates its widespread adoption e.g. in epidemiology, as criticized in \citep{keiding2016}.}

Another reason is that in Rubin's models, it is difficult to follow Rubin's call for `investigators [to] carefully describe their sample of trials and the ways in which they may differ from those in the target population' \citep[p. 698]{rubin1974}, given the difficulty to connect such a target population with the observed data in the formalism. 
The easiest setting would be to assume that the target population is sampled i.i.d. from the same distribution as the observed population.
In this case, one can draw connections between them through the conjured distribution, using finite sample theory to estimate quantities of interest and possible errors or confidence intervals.
This is, however, a very strong assumption; to say that the target population has a somewhat different generative process would mean to conjure a second abstract distribution that is somehow related to the first, from which the target population is then sampled.
Making the relations between them explicit is difficult since none of the relations observed sample -- first distribution -- second distribution -- target population is observable and no assumptions about them can be tested.

Given the observed lack of engagement with external validity, some researchers have recently started to argue for more explicit modelling of the target population.
For example, while Rubin acknowledges the vagueness of his notion of `representative' through quotation marks in the long quote above,  \citet[p. 4]{rudolph2023} argue that `[a]ll statements regarding representativeness should make clear the way in which the study results generalise, the target population the results are being generalised to, and the assumptions that must hold for that generalisation to be scientifically or statistically justifiable’. 
Similar points have also been made by \citet{westreich2019} and \citet{fox2022}.
In the following, we propose an amendment to RCMs which allows to directly model the target population and allows to connect it to observed data without any detour through abstract distributions.
This incentivises the engagement with concrete conditions that allow generalization and makes assumptions explicit and, retrospectively, testable.
Still, the proximity and direct relation to RCMs makes it possible to keep trained intuitions and methodologies.



\section{New framework}
\label{s:new_frame}




\subsection{Setup and notation}
\label{ss:setup}

We now introduce the setup and notation of the new version of the framework; it is close to the notation in \citep{manski2004}.
By $\XX$, $\YY$, and $\TT$ we denote the sets of possible covariates, outcomes (in $\RR$), and treatments, respectively.
We only consider binary treatments $\TT = \{0,1\}$ in the main text for easier comparison with RCMs, although nothing changes for our framework with multiple treatments
We assume that we have datapoints $(x_i, y_i, t_i)_{i \in \JJ}$  where $\JJ$ serves as the index set of our training data.
That is, for each unit $i \in \JJ$ with covariates $x_i \in \XX$, we have observed outcome $y_i \in \YY$ under treatment $t_i \in \TT$.\footnote{This means that there is no notation for counterfactual outcomes of observed datapoints.}

We then consider a target population $\II$ with an unknown \textbf{outcome function}\footnote{\citet{manski2004} calls this the `response function'.} 
\begin{equation}
    \y : \II \times \TT \to \YY
\end{equation}
such that $\y(i,t)$ denotes the outcome when treatment $t \in \TT$ is assigned to individual $i \in \II$.\footnote{The common SUTVA assumption precluding interactions between treatment assignments to different individuals is encoded in the fact that the outcome only depends on the individual's treatment. One could allow such interactions by taking as inputs $i$ and a treatment vector of length $|\II|$.}
Many methods for non-experimental settings require covariates; we denote  covariates of target units by $\x(i), i \in \II$ using a \textbf{representation function} 
\begin{equation}
    \x: \II \to \XX.
\end{equation}
Using a function for this highlights both that the covaruates of the target population may yet be unknown and that representing an individual $i$ through some covariates in a space $\XX$ is itself a deliberate action.

As mentioned above, the most common quantity of interest is the average treatment effect (ATE); the finite-population version in our setting would be 
\begin{equation}\label{eq:ATE}
    \frac{1}{|\II|} \sum_{i \in \II} \y(i, 1)
    - 
    \frac{1}{|\II|} \sum_{i \in \II} \y(i, 0).
\end{equation}
%
This theoretical quantity is the difference between the average outcomes of assigning either treatment or control to everyone\footnote{We generalise this to more complex treatment rules in Section~\ref{ss:treatment_rules}.}, the \textbf{average potential outcome (APO)} 
\begin{equation}\label{eq:APO}
    \avgIIyt := \frac{1}{|\II|} \sum_{i \in \II} \y(i, t).
\end{equation}
%

Our framework frames assumptions and results in terms of observable quantities that can be directly compared.
To this end, we introduce a shorthand notation for approximate equality:
\begin{thm}
\begin{definition}
    For $\eps > 0$, we say that two values $r,s \in \RR$ are $\eps$-similar if $|r - s| < \eps$. We write this as
    \begin{equation}
        r \approx_{\eps} s.
    \end{equation}
\end{definition}
\end{thm}

In the next section, we discuss fundamental assumptions that allow us to make predictions about the target population based on observed data.


\subsection{Experimental setting: RCTs}
\label{ss:RCTs}

In the comparatively simple case of RCTs, we do not need covariates for predicting the APO.
Here, we individuals are randomly assigned into treatment and control group. 
In the earlier example, this could mean that a lottery is used to decide who is offered job training in a population of unemployed people.
We denote treatment ($t=1$) and control ($t=0$) group in the observed data $\JJ$ by $\JJ_1$ and $\JJ_0$, that is,
\begin{equation}\label{eq:J_t}
    \JJ_t := \{i \in \JJ : t_i = t\}, \quad t \in \TT = \{0,1\}.
\end{equation}
The random assignment in RCTs can be used to justify the assumption that the average outcome in $\JJ_t$ approximates our quantity of interest  (\ref{eq:APO}), that is, the average outcome in the target population if we assign treatment $t$ to everyone:
\begin{equation}\label{eq:RCT-APO}
    \avgIIyt := \frac{1}{|\II|} \sum_{i \in \II} \y(i,t) \approx \frac{1}{|\JJ_t|} \sum_{i \in \JJ_t} y_i. 
\end{equation}
For this, we do not need to invoke any true distribution;
it is enough to assume that the partition into observed and target samples as well as the partition into control and treatment groups can be considered random.
That is, if the partition into $\II$ and $\JJ$ is random and the partition of $\JJ$ into $\JJ_0$ and $\JJ_1$ is random, then $\JJ_t$ under treatment $t$ is representative\footnote{\citet{rudolph2023} argue for such a notion of representativeness that is tied to a target population.} of $\II$ under treatment $t$: 
This means we treat them as being drawn from the same urn, similar to Neyman's original work.\footnote{It has been argued that such an urn model `applies rather neatly to the as-if randomized natural experiments of the social and health sciences' \citep[692]{freedman2006}.}
This justifies (\ref{eq:RCT-APO}) because (for large enough data sets) the vast majority of possible partitions will lead to roughly equal averages in both parts.
This can be shown with a Hoeffding inequality for finite samples (as in Proposition 1.2 of \citep{bardenet2015}), giving statements of the form `for 95\% of partitions, the difference between means is below 0.05'.

Such a measure on the number of admissible partitions can be made into a probabilistic statement (`with 95\% probability...') if we additionally assume that all partitions are equally which is made in RCMs in the form of the i.i.d. assumption and in Bayesian frameworks such as \citep{dawid2021} in the form of exchangeability.
In a finite population framework, this can, however, be neatly generalised to allow biased sampling schemes \citep{meng2018, meng2022}; we elaborate on the connection to non-probability sampling in Appendix~\ref{app:non_prob}.
Our approach thus highlights this equi-probability assumption and allows to easily incorporate deviations, in the form of correlations between outcome and group membership.
In our running example, this could take the form of incorporating e.g. fluctuations in general unemployment if the data collection happened a few years earlier.


\subsection{Predictions and assumptions}
\label{ss:predictors}

Causal inference becomes more complicated outside controlled experiments and most literature is concerned with observational or quasi-experimental settings.
For example, the data we have about the efficacy of job trainings typically does not come from RCTs.
In such settings, we incorporate additional information in the form of covariates $x \in \XX$.
As we will demonstrate, these covariates are used not to analyse the difference between treatment and control groups, but between the treatment/control parts of the observed sample on the one hand and the full target sample on the other.

\begin{figure}
    \centering
    \includegraphics[width=0.95\linewidth]{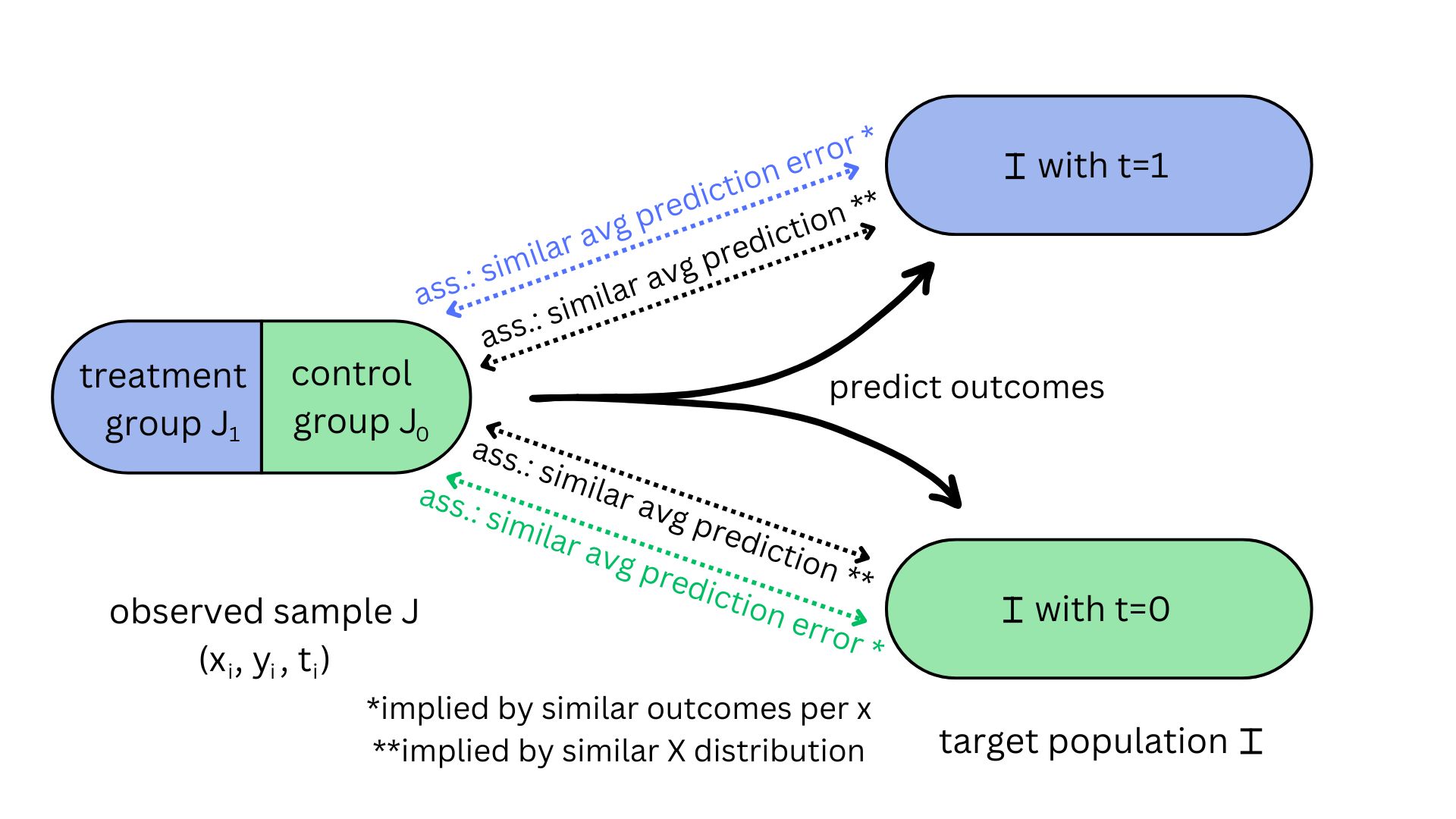}
    \caption{A schematic drawing of our framework: the observed sample is directly compared to the target sample, via an assumption regarding a similar distribution of covariates in $\JJ$ and $\II$ ($\eps$-SAP) and an assumption that a predictor based on $\JJ_t$ is also calibrated on $\II$ under treatment $t$ ($\delta$-CTP).}
    \label{fig:ours}
\end{figure}

The last important concept in our framework is a predictor
\begin{equation}
    p : \XX \times \TT \to \RR.
\end{equation}
As we will show now and, more extensively, in Appendix~\ref{app:new_persp}, different estimators can be characterised by different predictors $p$, all with the goal of estimating the APO on the target sample though the average prediction on the observed sample.
To make this work, two inductive assumptions are needed that tie the observed sample (and its treatment-wise groups) to the target sample (Figure~\ref{fig:ours}).
This dissolves the traditional separation into internal and external validity (which should be seen as an advantage, as we will argue in the next section).

For external validity, the standard framework typically assumes that the target data look like the observed data in the sense that they are sampled from the same marginal distribution of $X_i$; we require a more specific and testable property:
We assume that the average prediction on the observed population indexed by $\JJ$ is roughly the same as on the target population indexed by $\II$.
We formalise this for $\eps > 0$ as the \textbf{$\eps$-stable average predictions ($\eps$-SAP)} assumption
\begin{equation}\label{eq:SPA}
    \avgIIpt \approx_{\eps} \avgJJpt.
\end{equation}
where we denote the average prediction on observed and target sample by
\begin{equation*}
    \avgJJpt := \frac{1}{|\JJ|} \sum_{i \in \JJ} p(x_i, t)
\end{equation*}
and
\begin{equation*}
    \avgIIpt := \frac{1}{|\II|} \sum_{i \in \II} p(\x(i), t),
\end{equation*}
respectively.
While the assumption explicitly relates to the predictor $p$, it follows from the conventional assumption that covariates in sample and target population are sampled from the same distribution, at least for bounded $p$ and for large enough populations (see Section~\ref{ss:formal_comp}).

%

In addition to $\eps$-SAP, we need our predictor to work well on the target population.
This assumption is formalised as \textbf{$\delta$-calibration on the target population ($\delta$-CTP)},
\begin{equation}\label{eq:CTP}
    \avgIIyt \approx_\delta \avgIIpt.
\end{equation}
Essentially, this says that the errors of our predictor, when applied to the target population, will roughly average out.
Together, these two assumptions allow us to make an $\eps+\delta$-good approximation of the APO under treatment $t$:
\begin{equation}
    \avgIIyt
    \approx_\delta \avgIIpt
    \approx_{\eps} \avgJJpt.
\end{equation}
Different approaches to causal inference provide different estimators of the APO $\avgIIyt$ through different predictors $p$, for which the $\delta$-CTP assumption needs to be justified individually.
Typically, $p$ is calibrated on $\JJ_t$ for all $t \in \TT$ by design, so that $\delta$-CTP follows if we assume that the calibration error of $p$ on $\II$ is $\delta$-close to the calibration error on $\JJ_t$:
\begin{equation}
    \avgIIpt - \avgIIyt \approx_\delta 
    \sum_{i \in \JJ_t} p(x_i, t_i) - \sum_{i \in \JJ_t} y_i = 0.
\end{equation}

For example, we can see RCT-based inference as a using the degenerate predictor that predicts the group-wise mean for each individual, i.e. 
\begin{equation}\label{eq:p_rct}
    p: (x,t) \mapsto \frac{1}{|\JJ_t|} \sum_{i \in \JJ_t} y_t.
\end{equation}
Then 
\begin{equation}
    \avgIIpt = p(x,t) = \frac{1}{|\JJ_t|} \sum_{i \in \JJ_t} y_t
\end{equation}
s.t. $\delta$-CTP then boils down to the assumption that the mean outcome in $\JJ_t$ is $\delta$-close to the mean outcome in $\II$ under treatment $t$, which we have discussed above.
A more interesting case is matching or inverse probability weighting.


\subsection{Observational setting and matching}
\label{ss:matching}

One of the most basic techniques is \textbf{exact matching} \citep{rosenbaum1983}.
To capture this, define subgroups $\II^x, \JJ^x, \JJ_t^x$ for $x \in \XX, t \in \TT$,
\begin{align}
    \II^x &:= \{i \in \II\ :\ \x(i)=x\}\\
    \JJ^x &:= \{i \in \JJ\ :\ x_i=x\}\\
    \JJ_t^x &:= \{i \in \JJ_t\ :\ x_i=x\}.
\end{align}
For a sufficiently coarse-grained set of covariates, we may observe all combinations  $(x,t)$ of covariates and treatments -- which is often called `positivity' or `common support':
\begin{equation}\label{eq:comm_supp}
    \forall x \in \XX, t \in \TT: \JJ_t^x \neq \emptyset.
\end{equation}
Exact matching then uses the point-wise average predictor 
\begin{equation}\label{eq:p_matching}
    p: (x,t) \mapsto \frac{1}{|\JJ_t^x|} \sum_{i \in \JJ_t^x} y_i.
\end{equation}
This predictor is much more fine-grained that the RCT predictor (\ref{eq:p_rct}), as $p(x,t)$ predicts the observed $x$\emph{-wise} average outcome in group $\JJ_t$, rather than averaging over all of $\JJ_t$.
As we show in Proposition~\ref{prop:exact_match}, this predictor satisfies $\delta$-CTP (\ref{eq:CTP}) if the \textbf{average (signed) difference} between the $x$-wise mean outcomes
\begin{equation*}
    \avgIIytx := \frac{1}{|\II^x|} \sum_{i \in \II^x} \y(i,t)
    \quad \text{and} \quad 
    \avgJJytx := \frac{1}{|\JJ_t^x|} \sum_{i \in \JJ_t^x} y_i,
\end{equation*}
is not strongly biased above or below zero, that is,
\begin{equation}\label{eq:exact_match_avg_signed_diff}
    \left|  \sum_{x \in \XX} \frac{|\II^x|}{|\II|} \left( \avgIIytx - \avgJJytx \right) \right|
    < \delta.
\end{equation}
Note that in the limit of infinite data drawn from some distribution, the common unconfoundedness assumption (\ref{eq:unconf}) would imply that the term $\avgIIytx - \avgJJytx$ goes to zero \textit{for every} $x$ with probability one; 
this is strictly stronger than (\ref{eq:exact_match_avg_signed_diff}), as the latter allows that the differences for different $x$ values cancel each other out (see Section~\ref{ss:formal_comp}).
We now show that in practice, (\ref{eq:exact_match_avg_signed_diff}) is indeed sufficient for the exact matching predictor; using the exact matching predictor for predicting the APO then amounts to inverse probability weighting:
%
\begin{thm}
\begin{proposition}[Predicting the APO through matching] \label{prop:exact_match}
    \ \\
    Fix any $t \in \TT$.
    Assuming $\eps$-SAP (\ref{eq:SPA}) for $p$ as in (\ref{eq:p_matching}) and average (signed) difference below $\delta$ as in (\ref{eq:exact_match_avg_signed_diff}), the exact matching predictor (\ref{eq:p_matching}) gives us an $(\epsilon+\delta)$-good approximation of the APO $\avgIIyt$:
    \begin{equation}
        \left| \avgIIyt
        - \frac{1}{|\JJ|} \sum_{i \in \JJ_t} \frac{y_i}{\prop_t(x_i)} \right| 
        < \epsilon + \delta,
    \end{equation}
    where $\prop_t(x) := \frac{|\JJ_t^x|}{|\JJ^x|}$ is the observed propensity score for treatment $t$.
\end{proposition}
\end{thm}
\begin{proof} 
    First, we get $\delta$-CTP for predictor from (\ref{eq:exact_match_avg_signed_diff}) via
    \begin{align}
        \avgIIyt - \avgIIpt
        &= \frac{1}{|\II|} \sum_{i \in \II} \y(i, t) - p(\x(i),t) \\
        &= \frac{1}{|\II|} \sum_{x \in \XX} \sum_{i \in \II^x}  \left(\y(i,t) - \sum_{j \in \JJ_t^x} \frac{y_j}{|\JJ_t^x|} \right) \\
        &= \frac{1}{|\II|} \sum_{x \in \XX} |\II^x| \left( \sum_{i \in \II^x} \frac{\y(i,t)}{|\II^x|} - \sum_{i \in \JJ_t^x} \frac{y_i}{|\JJ_t^x|} \right)\\
        &\approx_\delta 0.
    \end{align}
    Then
    \begin{align}
            \frac{1}{|\II|} \sum_{i \in \II} \y(i, t) 
            &\approx_\delta \frac{1}{|\II|} \sum_{i \in \II} p(\x(i),t) \\
            &\approx_\eps \frac{1}{|\JJ|} \sum_{i \in \JJ} p(x_i,t) \label{eq:population}\\
            &= \frac{1}{|\JJ|} \sum_{x \in \XX} |\JJ^x| \cdot p(x,t)\\
            &= \frac{1}{|\JJ|} \sum_{x \in \XX} \frac{|\JJ^x|}{|\JJ_t^x|} \sum_{i \in \JJ_t^x} y_i\\
            &= \frac{1}{|\JJ|} \sum_{i \in \JJ_t} \frac{y_i}{\prop_t(x_i)},
    \end{align}
    where (\ref{eq:population}) uses $\eps$-SAP (\ref{eq:SPA}).
\end{proof}


When the common support assumption (\ref{eq:comm_supp}) is not reasonable, one may be able to use a more coarse-grained approach.
Using `coarsened exact matching', we can predict averages not on every $x \in \XX$ but on suitable subsets $U \in \Xpart$ where $\Xpart$ is a partition of $\XX$.
We discuss this, along with other methods such as doubly robust estimators, instrumental variables, and diff-in-diff, in Appendix~\ref{app:new_persp}.


\section{Comparing the frameworks}
\label{s:compare}

In this section, we compare the proposed framework with RCMs.
We start with a descriptive comparison in mathematical form, relating the assumptions of $\eps$-SAP and $\delta$-CTP with i.i.d. sampling and unconfoundedness (Section~\ref{ss:formal_comp}).
After that, we give a more subjective account of the practical advantages we see in the new framework (Section~\ref{ss:practical_comp}).


\subsection{Formal considerations}
\label{ss:formal_comp}

We first note that $\eps$-SAP follows from the conventional assumption that covariates in past and future are sampled from the same distribution, at least for bounded $p$ and for large enough populations:
\begin{thm}
\begin{remark}[average prediction in RCMs] \label{rem:sap}
    \ \\
    For datapoints $x_1, ..., x_N$ sampled i.i.d. from a distribution $P$ that governs $X$ on $\XX$ and some bounded predictor $p : \XX \times \TT \to \RR$, it follows that $\forall t \in \TT$ 
    \begin{equation}
        \lim_{N \to \infty} \frac{1}{N} \sum_{i=1}^N p(x_i, t) = \EE_P[p(X,t)]
    \end{equation}
    almost surely. 
\end{remark}
\end{thm}
This means that if both datasets $\JJ$ and $\II$ are assumed to be sampled i.i.d. from the same distribution over $\XX$, this implies that (\ref{eq:SPA}) holds almost surely in the limit of infinite data for any $\eps>0$.

We can also relate the $\delta$-CTP condition to the RCM framework; this can take different forms, as discussed in the context of different predictors, but often relies on the common unconfoundedness assumption,
\begin{equation}\label{eq:unconf2}
    Y_{ti} \indep T_i\ |\ X_i.
\end{equation}
It is often suggested that the unconfoundedness setting is ``probably the most important one in practice in the modern CI literature'' \citep[p. 1163]{imbens2020}.
Judea Pearl complains that  ``I have yet to find a single person who can explain what [it] means in a language spoken by those who need to make this assumption or assess its plausibility in a given problem'' \citep[p.281]{pearl2018}.
Guido \citet{imbens2020} cites this passage and shoots back that ``simply assuming that one knows or can consistently estimate the joint distribution of all variables in the model'' is also ``not helpful'' (p. 1154).
Imbens later adds that unconfoundedness ``is so common and well studied that merely referring to its label is probably sufficient for researchers to understand what is being assumed \citep[p. 1164]{imbens2020}.
To what extent it is well understood is not easy to settle, but it seems clear that ``the unconfoundedness assumption is not directly testable'' \citep[p. 17]{imbens2024}.
What our framework provides is an inductive assumption that can be tested in hindsight, which takes different forms for different predictors (see Section \ref{s:new_frame} and Appendix~\ref{app:new_persp}) and can be derived from the high-level idea unconfoundedness:

%
\begin{thm}
\begin{remark}[conditional means under unconfoundedness] \label{rem:ase}
    \ \\ 
    For a distribution $P$ over $\XX \times \YY_t \times \TT$ satisfying unconfoundedness (\ref{eq:unconf2}) and datapoints $(y_1, t_1), ..., (y_N,t_N)$ sampled i.i.d. from the conditional distribution $P(Y_t,T|x)$, we have, for $\JJ_t^x(N):=\{1 \leq i \leq N : t_i = t\}$,
    \begin{equation}
        \lim_{N \to \infty} \frac{1}{|\JJ_t^x(N)|} \sum_{i \in \JJ_t^x(N)} y_i\ 
        =\ \EE_P[Y | X=x,T=t] 
    \end{equation}
    almost surely.
    This means that $\avgJJytx$ converges to the conditional expectation for each $x$ and $t$, which is also the limit to which $\avgIIytx$ converges (trivially).
\end{remark}
\end{thm}
Thus, for $|\JJ_t^x|, |\II^x| \to \infty$, $\avgIIytx - \avgJJytx$ goes to zero such that (\ref{eq:exact_match_avg_signed_diff}) is easily satisfied (for any $\delta$) with enough data.
Note that this is indeed much stronger than (\ref{eq:exact_match_avg_signed_diff}) since the latter only requires that the signed average of the differences between the conditional means is small, whereas the remark implies that all absolute differences go to zero.
Now (\ref{eq:exact_match_avg_signed_diff}) also implies that $\delta$-CTP is satisfied for the exact matching predictor, which justifies inverse probability weighting (Proposition~\ref{prop:exact_match}).


%

While uncounfoundedness is not testable, one can analyse to what extent the validity of results are sensitive to the violation of unconfoundedness in the form of unobserved confounders.
In such sensitivity analysis, it is often assumed that outcomes depend on the unobserved confounders via some specified functional form -- particularly common is linearly dependence, since sensitivity analysis is mostly developed for linear regression models. 
We now show that this is also doable for our framework, and in particular includes the modelling of generalisation to target populations:
In the simplest case of assuming linear dependence on unobserved confounders, we may formalise this by adding a term $u \cdot \gamma$ to the $\delta$-CTP:\footnote{This is not too different to the data deficiency coefficient applied to model errors as done in \citep{meng2022} (with $r$ seen as a correlation) -- for some further observations on connections to the non-probability sampling literature, see Appendix~\ref{app:non_prob}.} 
\begin{thm}
\begin{remark}[sensitivity to unobserved confounders] \label{rem:sensitiv}
    \ \\ 
    If we relax the $\delta$-CTP assumption by assuming 
    \begin{equation}
        \avgIIyt \approx_\delta \avgJJpt + u \cdot \gamma
    \end{equation}
    for some unobserved aggregate quantity $u$ that may change between $\JJ_t$ and $\II$ by some value of $r \in R \subset \RR$ and $\gamma \in \Gamma \subset \RR$, we can bound the APO by
    \begin{equation}
        \avgIIyt \geq 
        \avgJJpt - (\eps + \delta) + \min_{\Gamma, R} \gamma \cdot r,
    \end{equation}
    and
    \begin{equation}
        \avgIIyt \leq 
        \avgJJpt + (\eps + \delta) + \max_{\Gamma, R} \gamma \cdot r.
    \end{equation}
\end{remark}
\end{thm}
%
Note that this is not too different from the idea of e-values \citep{vanderweele2017} for sensitivity analysis, the main difference being our focus on summation rather than ratios; indeed, if we were interested in looser bounds based on maximising over coefficients per-strata $x$ (as for e-values \cite{ding2016}), we could drop the assumption of a global coefficient $\gamma$ and instead allow a different coefficient for each $x$.


\subsection{Practical considerations}
\label{ss:practical_comp}

While sensitivity analyses can be useful, they still do not make the fundamental assumption of (approximate) unconfoundedness testable.
Placebo tests go a bit further but rely on further untestable assumptions.
A central aspect of the new framework is, then, that inductive assumptions can be formulated in ways that are directly testable and relatable to predictive success -- while still allowing to use intuitions in terms of unconfoundedness to argue for the plausibility of the more concrete assumptions.
Another assumption that is easily overlooked is the assumption of (i.i.d.) sampling which, as a relation between observables and a distribution, is difficult to even formalise. 
While the problem can be seen as less problematic for the pursuit of internal validity (as the distributions is per definition defined for the observed sample), this becomes more problematic when the question of generalisation or transferability to a target population.
The new framework avoids this concept altogether (though it can be invoked as a limiting case, as in the related literature on non-probability sampling).

Perhaps most fundamental difference, however, is the shifted aim: from identification to prediction.\footnote{This view has predecessors even for the case of programme evaluation; e.g. \citet[p. 184]{berk1987} notes that
`evaluations of program impact necessarily involve predictions' which `involves expectations about the likely result under two or more conditions; they are predictions of the what-if variety.’}
The notion of identification hinges on the notion of unobserved true parameters.
Such parameters do not exist in our framework.
This difference also has ramifications for the direct modelling of target populations, testability, the divide between internal and external validity.
The lack of testability for assumptions like unconfoundedness has been discussed above.
Our proposal is in the spirit of \citet[p. 33]{freedman1995} suggesting that `[u]sing models to make predictions of the future, or the results of interventions, would be a valuable corrective', but goes even further by specifying testable assumptions, which makes it possible to distinguish between potential sources of error and making predictions the focus of modelling.
This also entails that the target population is explicitly included in the model, which requires (and allows) the modeller to directly grapple with the question of `external validity'.\footnote{We are not claiming that it is impossible in principle to capture generalisation with conventional frameworks. For a review of some recent attempts (and a call for more interest in the problem) see \citet{findley2021}.}

Another marked difference in our framework is that the separation between internal and external validity dissolves.
This is a direct implication of abandoning the idea of identifying supposed true parameters of abstract generative distributions.
The concepts of external and internal validity are so engrained in social science research that this may seem extremely counter-intuitive to the experienced researcher.
But also this idea is not without precedence: Indeed, the separate consideration of internal and external validity has recently been criticised as a reason for the neglect of external validity (Section~\ref{ss:target}) and the ensuing negative impact on public health research \citep{westreich2019}.
Furthermore, one can recover a version of internal validity by taking the sample population as the target population; we discuss this in the context of difference-in-difference estimators in Appendix~\ref{ss:d-in-d}.

Another, minor, novel aspect of our framework is that now the APO is the undisputed focus of the statistical enterprise, instead if the ATE.
While some estimators explicitly estimate the APOs as a first step, they still tend to be understood as ATE estimators.
Our framework more explicitly takes the APOs as the fundamental quantities, the ATE is not more than a formal comparison between APOs.


A more high-level difference is that the new framework explicitly works on averages rather than individuals.
In RCMs, ATEs are typically seen as averages of individual treatment effects -- indeed, the subscript $i$ attached to all variables is meant to convey that the parameters and functions directly apply to each individual.
Combined with high hopes for Machine Learning techniques (which use a similar formalism), there is an increasing push to `fully personalized treatment effect estimates’ \citep[p. 7353]{athey2016}.
This is despite the nature of statistics as a field capturing aggregate behaviour as well as cautionary voices pointing out the complexity of the social world (Section~\ref{ss:distribution}), where noise cannot be neatly controlled as in physical laboratories \citep[p. 187]{berk1987}.\footnote{Sander Greenland calls `soft sciences' those that cannot expect to discover numerically precise and general contextual laws analogous to those in physics' \citep[p. 4]{greenland2017}.}
In contrast, our framework dispenses not only with the notion individual effects but also discards the aim of estimating individual quantities; `true' conditional distributions are not defined and the goal is specified as predicting aggregate properties of outcomes in the target population.
So far, we have focused exclusively on the APO, i.e. the average, but other aggregate properties are possible, as discussed in the next section.





\section{Beyond APO and ATE}
\label{s:beyond_apo}

So far, we have only considered APOs, that is, the \emph{average} outcome if the \emph{same} treatment is applied to everyone.
This is enough to predict the average treatment effect, which is often taken to be the goal of causal inference.
There are, however, cases where this is not what we are interested in.
We, in turn, discuss relaxations of `same' and of `average'.

\subsection{Personalised aka conditional treatment rules}
\label{ss:treatment_rules}

Sometimes, we are not interested in applying the same treatment to everyone but instead want make treatment conditional on observed attributes.
In this case, we may like to predict the average outcome of more complex covariate-based treatment rules $\pi : \XX \to \TT$, 
\begin{equation}
    \frac{1}{|\II|} \sum_{i \in \II} \y(i, \pi(\x(i))).
\end{equation}
In this general formulation, assigning the same treatment rule to everyone, as considered so far, is captured by the degenerate policies $\pi_t : x \mapsto t$ for $t \in \TT$.
In recent years, starting with \citep{manski2004}, there has been an increasing focus on learning more sophisticated treatment rules based on data.
While we do not discuss the learning part in this paper, we note that our framework straightforwardly applies to predicting the average outcomes of such treatment rules.\footnote{They are sometimes called  `individualised' treatment rules -- `conditional' is arguably a better descriptor, as they are conditional on the attributes taken into account (which is a modelling choice), rather than tailored to specific individuals.}

To apply our analysis to such treatment rules $\pi : \XX \to \TT$, it suffices to consider the sub-populations induced by the level sets of $\pi$, that is, 
\begin{equation}
    \XX_t := \pi^{\textbf{-}1}(t) = \{x \in \XX : \pi(x) = t\}
\end{equation}
for $t \in \TT$.
Based on this, we partition $\JJ$ and $\II$ into subsets 
\begin{equation}
    \JJ^{\XX_t} := \{i \in \JJ : x_i \in \XX_t\}
    \quad \text{and} \quad
    \II^{\XX_t} := \{i \in \II : \x(i) \in \XX_t\},
\end{equation}
then apply our machinery to all the pairs $\JJ^{\XX_t}, \II^{\XX_t}$ instead of $\JJ, \II$.
For binary treatment $\TT = \{0,1\}$, this only means we consider two sub-populations instead of one population.
The assumptions connecting $\JJ^{\XX_t}$ and $\II^{\XX_t}$ for each $t$ are then analogous to those that we used in this paper to connect $\JJ$ and $\II$.
For example, the assumption that two human populations $\JJ$ and $\II$ are similar in all relevant respects is hardly weaker than assuming that the respective sub-populations of (for example) people above 40  are similar, as are those below 40.
However, this becomes stronger and stronger if we require this for more and more policies $\pi$ simultaneously; requiring this for all possible policies would mean that the two populations must be exactly alike.\footnote{For many other problems such as structural risk minimisation \citep{vapnik1982}, multi-calibration \citep{hebert2018}, and randomness \citep{vonMises1964}, a similar necessity for restricting the number of considered partitions has been observed; see also \citep{derr2022}.}

Beyond deterministic treatment rules $\pi : \XX \to \TT$, one may also be interested in more general \emph{stochastic treatment rules} $\pi : \XX \to \Delta(\TT)$, where $\Delta(\TT)$ denotes the set of probability distributions over $\TT$.
For binary treatment decisions, that means that $\pi$ assigns a treatment probability to each $x \in \XX$.
There are at least two distinct arguments for using such stochastic treatment rules: an ethical, and an epistemic one.\footnote{\citet{jain2024} have recently also advocated stochastic allocation rules.}
The ethical argument is that hard cut-offs are unfair because two people on opposite sides of the threshold are treated very differently \citep{vredenburgh2022}.
The epistemic argument is that we often do not have much data for some covariates, and if we then never assign some treatments to them, we will never gain that information.
This can exacerbate inequality in the case of underrepresented groups, as discussed in \citep{oneil2017} and analysed in a bandit setting in \citep{li2020}.
In causal inference, this resurfaces in terms of the assumptions we rely on when analysing data.
If we use stochastic treatment rules, we get a well-defined propensity score by design that we can use for subsequent modelling.
That is, we satisfy the assumption of `missing at random' \citep{rubin1976} (see Footnote~\ref{fn:missing-at-random}) \textit{and} get direct access to the conditional probabilities.
If the propensity score never reaches 0 or 100 per cent, we also satisfy positivity/common support.
We have suggested calibration on sets of equal propensity as a testable criterion.
This suggests that it could be beneficial to use treatment rules which only assign a limited set of treatment probabilities, such as $\{0.1, 0.3, 0.5, 0.7, 0.9\}$, instead of the whole interval $[0,1]$, in order to facilitate the analysis.

\subsection{Beyond averages}
\label{ss:beyond_average}


In general, a social planner is interested in choosing the policy that maximises desirable properties of  the distributions of outcomes.
This gives a further reason to focus on treatment-wise potential outcomes rather than supposed treatment effects: 
Even if meaningful, the distribution of individual treatment effects would not allow us to infer other properties of the outcome distributions beyond the mean \citep[714]{manski1996}.
For simplicity and to directly compare with the bulk of the RCM literature, we have so far restricted ourselves to averages 
-- but we do consider it important to go beyond this.
In the following, we consider three alternatives to the APO, that is, other properties of the potential outcome distribution.
This means we consider scalar predictions rather than probabilities for binary outcomes, since for binary outcomes, the average would specify the complete distribution.
An example of such an outcome is income.

A first quantity of interest is the proportion of individuals with an income above a certain threshold (such as the poverty line).
This just collapses into the APO if we consider the binary outcome of whether an individual has an income above that threshold.
Therefore, we do not discuss this property further.

A second potentially interesting property of the outcome distribution is the median or, more generally, quantiles.
The target quantity is the $p$-quantile of the (empirical) target distribution for $p \in (0,1)$, that is, the inverse $F_{\II,t}^{-1}(p)$ of the cumulative distribution function
\begin{equation}
    F_{\II,t}(y) := \frac{|\{i \in \II : \y(i,t) \leq y\}|}{|\II|}.
\end{equation}
This is not necessarily well-defined so we make it more specific by defining 
\begin{equation}
    \qtp := \min \left\{ y \in \YY :  F_{\II,t}(y) \geq p \right\}.
\end{equation}
In words, $\qtp$ is the lowest outcome threshold such that at least $p\%$ of the target population fall below it.
It turns out that this problem can also be almost reduced to that of APOs -- by fixing the value of the quantile $\qtp$ and then again considering the binary outcome of whether an individual has an income above that threshold.
There are two caveats to this reduction.
The first one is that we do not see the quantile $\qtp$, so we need to consider the estimator of that quantile as our threshold.
Still, by the above reasoning, we can argue that the proportion of people above the chosen threshold should remain similar.
The second caveat is that small variation in the proportion may correspond to a large variation in the quantile if the differences between the outcomes around the threshold are high.
This requires a further assumption, bound this variation by requiring that for some function $\alpha: \RR \to \RR$, $\forall y \in \{\y(i,t) : i \in \II\}$,
\begin{equation}\label{eq:local_lipschitz}
    | F_{\II,t}(y) - p | < \beta
    \quad \rightarrow \quad 
    |y - \qtpHat | < \alpha(\beta).
\end{equation}
While it is then analogous to the case of average outcomes (modulo the mentioned changes), we walk through the quantile case in Appendix~\ref{app:quantiles} as these changes may not be as intuitive.

A third and last type of quantity, that we want to at least hint at, is based on the notion of social welfare functions (SWFs).
SWFs take a set of outcomes and reduce them to a number, such as the average, but they may also pay attention to other aspects of the distribution.
SWFs that is particularly interesting are rank-dependent and equality-minded \citep{kitagawa2021} and can be characterised as 
\begin{equation}
    W_\omega(\II, t)
    := \frac{1}{Z} \sum_{i=1}^N y_i \cdot \omega(F_{\II,t}(y_i)),
\end{equation}
where $Z := \sum_{i=1}^N \omega(F_{\II,t}(y_i))$ is a normalising constant and $\omega$ is non-negative and monotonically increasing, i.e. assigning lower weights to higher outcomes based on their rank/percentile.
The average corresponds to a constant $\omega$; other SWFs are the minimum, that is, how the worst off are doing, and measures in between.
From the more complex characterisation of such outcomes, it is already apparent that it may in general be more difficult to characterise as well as satisfy the required assumptions to predict this with precision.
While we consider this an important topic, we leave it for future work.



\section{Discussion}
\label{s:discussion}

In this paper, we have provided a mathematical framework that takes inspiration from Rubin's potential outcome framework but models the populations without relying on abstract distributions.
The focus in this framework shifts from identifying abstract parameters in stipulated distributions to predicting outcomes on the target population.
In practice, this allows to directly predict the observable effects of policies and, by relying only on testable assumptions, to retrospectively analyse sources of error in the modelling.
These advantages are particularly relevant for evaluating and informing concrete policies and interventions, which are often considered to be the most straightforward application setting for RCMs.


In conformity with Occam's razor, we also avoid unnecessary metaphysical assumptions in the form of well-defined counterfactuals, individual causal effects, or joint probability distributions.
We have already discussed the idealising assumption of well-defined probability distributions  from which the observations are sampled -- something for which in the complex systems modelled by social and health sciences, `you need a lot of good arguments' \citep[p.325]{cartwright1999}.
Our framework also suggests, like that of \citet{dawid2000, dawid2021}, `to reconfigure causal inference as the task of predicting what would happen under a hypothetical future intervention, on the basis of whatever (typically observational) data are available' \citep[p.299]{dawid2022}.
This is not as radical as it may seem, similar sentiments have been expressed, for example\footnote{Already Ragnar Frisch, `the founding father of modern econometric causal policy analysis' \citep[p. 4]{heckman2024} argued that `the scientific [...] problem of causality is essentially a problem regarding our way of thinking, not a problem regarding the nature of the exterior world' \citep[p. 36]{frisch1930}.}, by \citet{berk1987}, \citep{greenland2012}\footnote{Greenland here even believes to discern `a subtle conceptual revolution that recognizes causal inference as a prediction problem' (p. 44).} or \citet{hernan2016}, who notes that 
\begin{quote}
    `The goal of the potential outcomes framework is not to identify causes--or to ``prove causality'', as it sometimes said. That causality cannot be proven was already forcibly argued by Hume in the 18th century. Rather, quantitative counterfactual inference helps us predict what would happen under different interventions' \citep[p. 679]{hernan2016}.
\end{quote}
%
These more interpretational considerations should, however, not divert attention from the practical benefits of explicitly modelling the target population and making testable assumption.
Indeed, one may see the proposed framework either as a less metaphysically loaded alternative to RCMs or simply as an empirically minded amendment.
Although our contribution is only the first step in developing the new version of the framework, we hope to thereby contribute to a solid theoretical basis for using causal inference methods in practice.


\section*{Acknowledgements}
This work was funded by the Deutsche Forschungsgemeinschaft (DFG, German Research Foundation) under Germany’s Excellence Strategy—EXC number 2064/1—Project number 390727645 as well as the German Federal Ministry of Education and Research (BMBF): Tübingen AI Center, FKZ: 01IS18039A. 
The authors thank the International Max Planck Research School for Intelligent Systems (IMPRS-IS) for supporting Benedikt Höltgen.


\printbibliography


\appendix

\section{New perspectives on popular methods}
\label{app:new_persp}

Statisticians, economists, and others have developed various methods in the conventional RCM framework, many of which are well-established by now.
In this appendix, we survey a few of them and demonstrate how we can recover them in our framework with weaker assumptions.
The purpose of this is twofold:
First, to showcase the new framework and demonstrate how it can capture and explain established methods, and second, to inspect these methods themselves and provide new perspectives that enhance our understanding of them.


\subsection{Coarsened exact matching}
When the (strong) common support assumption is not reasonable, one may be able to use a more coarse-grained approach, that is, \textbf{coarsened exact matching}.
On this approach, we predict averages not on every $x \in \XX$ but on suitable subsets $U \in \Xpart$ where $\Xpart$ is a partition of $\XX$.
For $U \in \Xpart$, $t \in \TT$, define
\begin{align}
    \II^U &:= \{i \in \II : \x(i) \in U\} \\
    \JJ_t^U &:= \{i \in \JJ : x_i \in U \wedge t_i=t\}.
\end{align}
Then  we may use the predictor
\begin{equation}\label{eq:p_coarse_matching}
    p: (x,t) \mapsto \frac{1}{|\JJ_t^{U(x)}|} \sum_{i \in \JJ_t^{U(x)}} y_i,
\end{equation}
(with $U(x)$ denoting the $U \in \Pi$ containing $x$) to predict the APO:
Again, $\delta$-CFD can be expressed as a signed average difference, now with the differences in subsets $\II^U$:
\begin{equation}\label{eq:coarse_match_avg_signed_diff}
    \left| \sum_{x \in \XX} \frac{|\II^U|}{|\II|} \left( \avgIIytU - \avgJJytU \right) \right|
    < \delta.
\end{equation}
Note that this is also satisfied if the differences between the average outcomes are $\delta$-small for all $U$, i.e.
\begin{equation}\label{eq:causal_ass_subset}
    \forall U \in \Xpart: \quad
    \frac{1}{|\II^U|} \sum_{i \in \II^U} \y(i,s)
    \approx_{\delta} \frac{1}{|\JJ_t^U|} \sum_{i \in \JJ_t^U} y_i.
\end{equation}
Also note that $\eps$-SP for the coarsened matching predictor is satisfied if
\begin{equation}
    \forall U \in \Xpart: \left| \frac{|\II^U|}{|\II|} - \frac{|\JJ^U|}{|\JJ|} \right| < \frac{\eps \cdot |\JJ_t^{U}|}{\sum_{i \in \JJ_t^{U}} y_i}.
\end{equation}
%
\begin{thm}
    \begin{proposition}[Predicting the APO through coarsened matching]
        \ \\
        Fix any $t \in \TT$.
        Assuming $\eps$-SP (\ref{eq:SPA}) for $p$ as in (\ref{eq:p_coarse_matching}) and $\delta$-average signed difference as in (\ref{eq:coarse_match_avg_signed_diff}), the coarsened exact matching predictor gives us an $(\epsilon+\delta)$-good approximation of the APO for $t$:
        \begin{equation}
            \left| \avgIIyt
            - \frac{1}{|\JJ|} \sum_{i \in \JJ_t} \frac{y_i}{\proptU(x_i)} \right| 
            < \epsilon + \delta,
        \end{equation}
        where $\proptU(U) := \frac{|\JJ_t^U|}{|\JJ^U|}$ and $\proptU(x) := \proptU(U(x))$, with $U(x)$ being the $U \in \Xpart$ with $x \in U$.
    \end{proposition}
\end{thm}
\begin{proof}
    First, we get $\delta$-CFD from (\ref{eq:coarse_match_avg_signed_diff}) via
    \begin{align}
        \frac{1}{|\II|} \sum_{i \in \II} \y(i, t) - p(\x(i),t)
        &= \frac{1}{|\II|} \sum_{U \in \Xpart} \sum_{i: \x(i) \in U}  \left(\y(i,s) - \sum_{j \in \JJ_t^{U}} \frac{y_j}{|\JJ_t^{U}|} \right) \\
        &= \frac{1}{|\II|} \sum_{U \in \Xpart} |\II^U| \left( \sum_{i \in \II^U} \frac{\y(i,s)}{|\II^U|} - \sum_{i \in \JJ_t^{U}} \frac{y_i}{|\JJ_t^{U}|} \right) \\
        &\approx_\delta 0.
    \end{align}
    Then
    \begin{align}
        \avgIIyt = \frac{1}{|\II|} \sum_{i \in \II} \y(i, t) 
        &\approx_\delta \frac{1}{|\II|} \sum_{i \in \II} p(\x(i),t) \\
        &\approx_\eps \frac{1}{|\JJ|} \sum_{i \in \JJ} p(x_i,t)\\
        &= \frac{1}{|\JJ|} \sum_{U \in \Xpart} \sum_{i \in \JJ^{U}} p(x_i,t)\\
        &= \frac{1}{|\JJ|} \sum_{U \in \Xpart} |\JJ^U| \cdot \frac{1}{|\JJ_t^{U}|} \sum_{i \in \JJ_t^{U}} y_i\\
        &= \frac{1}{|\JJ|} \sum_{U \in \Xpart} \frac{1}{\proptU(U)} \sum_{i \in \JJ_t^U} y_i\\
        &= \frac{1}{|\JJ|} \sum_{i \in \JJ_t} \frac{y_i}{\proptU(x_i)}.
    \end{align}
\end{proof}

We can derive some insights from our analysis.
The RCM literature typically assumes that there is a true underlying local propensity score that we can estimate, which allows for the construction of consistent estimators of the ATE.
Outside of study designs that use weighted lotteries, it is not clear what the propensity score corresponds to in the real world; it is, thus, questionable if it makes sense to estimate this missing ground truth.
If assumptions such as smoothness of the propensity score in $\XX$ are made (which are implicit for any estimation method), it seems less problematic to explicitly assume `constant propensity' on all $U \in \Xpart$.
This would already imply (\ref{eq:causal_ass_subset}) (together with the common assumption that $\II$ and $\JJ$ are sampled from the same distribution).
Note that our analysis is very similar to the propensity score theorem \citep{rosenbaum1983} which derives $Y_{0i}, Y_{1i} \indep T_i | \prop(X_i)$ from the unconfoundedness assumption $Y_{0i}, Y_{1i} \indep T_i | X_i$:
In our derivation, we weaken the unconfoundedness assumption to the statement that future $t$-outcomes are on average similar to the data we have for $\JJ_t$ and then show that we can `condition on' sets of equal propensity.
This can be seen as interpolating between exact matching and RCTs, where we basically assume that the propensity score is constant everywhere.

We have adopted the name  `coarsened exact matching' from \citep{iacus2012}.
They propose to coarsen variable-wise, e.g. applying a grid; we have shown that if we coarsen to sets that can be thought to have a constant propensity score, the predictions are well-founded.
In line with this, it has also been suggested in \citep{little1986} to coarsen the considered covariates into groups of similar predicted propensity score to reduce variance, which they call `response propensity stratification'.
\citet{kang2007} report that this indeed provides more robust estimates.
As a last remark, one might say, in line with the general theme of our approach, that predictions based on matching approaches do not match treatment to control group, but both observed groups to anticipated future data.


\subsection{General calibrated predictors}

After having discussed specific matching predictors that can be expressed in terms of observed data, we now consider arbitrary $p: \XX \times \TT \to \RR$.
An example for a wider class of predictors is given by the increasingly used Machine Learning algorithms.\footnote{For now, we sidestep questions of overfitting by simply taking $\JJ$ to be a validation set.}
Compared to matching, we now have less reason to believe in a low average signed error because learning algorithms often lead to systematic errors through their inductive biases.
Analogous to coarsened matching, one may thus aim for low area-wise error on a partition $\Xpart$ of $\XX$ where we can hope that the error will be similar on future data, in the sense of
\begin{equation}\label{eq:ML_eq_prop}
    \frac{1}{|\II^U|} \sum_{i \in \II^U} p(\x(i),t) -\y(i, t)
    \approx_\delta \frac{1}{|\JJ_t^U|} \sum_{i \in \JJ_t^U} p(x_i,t) - y_i.
\end{equation}
%
Intuitively, this would be guaranteed (in the limit) on areas of `constant propensity':
In the distributional framework of RCMs, constant propensity on $U$ would mean that the distribution of $X$ on $\JJ_t^U$ is equal to that of $\JJ^U$ which is, in turn, equal to that on $\II^U$ -- which means that, since $P(Y|X)$ is the same on $\JJ_t$ as on $\II$ via the unconfoundedness assumption, the joint distribution $P(X,Y)$ is the same on $\JJ_t^U$ as on $\II^U$.
In settings where such distributions are difficult to justify, one may look for other ways to justify ($\ref{eq:ML_eq_prop}$), which is, after all, much weaker than assumptions about propensity scores.

\begin{thm}
\begin{proposition}[Predicting the APO through ML]
    \ \\
    Fix any $t \in \TT$.
    Assuming $\eps$-SP (\ref{eq:SPA}) for some $p$ and (\ref{eq:ML_eq_prop}) for all $U$ in some partition of $\XX$, we get an $(\epsilon+\delta)$-close approximation of the APO for $t$:
    \begin{equation}
        \left| \avgIIyt
        - \frac{1}{|\JJ|} \sum_{i \in \JJ} p(x_i,t) \right| 
        < \epsilon + \delta.
    \end{equation}
\end{proposition}
\end{thm}
\begin{proof}
    \begin{align}
        \avgIIyt = \frac{1}{|\II|} \sum_{i \in \II} \y(i, t) 
        &= \sum_{U \in \Xpart}  \frac{|\II^U|}{|\II|} \frac{1}{|\II^U|} \sum_{i \in \II^U} \y(i, t) \\
        &\approx_\delta \sum_{U \in \Xpart}  \frac{|\II^U|}{|\II|} \frac{1}{|\II^U|} \sum_{i \in \II^U} p(\x(i), t) \\
        &= \frac{1}{|\II|} \sum_{i \in \II} p(\x(i),t) \\
        &\approx_\eps \frac{1}{|\JJ|} \sum_{i \in \JJ} p(x_i,t).
    \end{align}
\end{proof}


\subsection{Doubly robust estimators}

As discussed above (see (\ref{eq:exact_match_avg_signed_diff})), the calibration error on future data can be expressed in terms of the average $x$-wise (signed) prediction error:
\begin{align}
    \avgIIpt - \avgIIyt
    &= \frac{1}{|\II|} \sum_{x \in \XX} \sum_{i \in \II^x}  \left( p(x,t) - \y(i,t) \right) \\
    &= \sum_{x \in \XX} \frac{|\II^x|}{|\II|} \left( p(x,t) - \avgIIytx \right).
\end{align}
%
For the matching predictors, one may hope that this is close to zero through something akin to the unconfoundedness assumption, given that the predictor is simply the local observed average outcome (see Remark~\ref{rem:ase}).

Alternatively, one may try to predict the local weights, in doubly robust estimators.
These estimators use predictions $p$ and $w$ of both the mean and the propensity score; they allow one to correctly predict the APO when only one of the two is correct.
%
In our framework, the doubly robust estimator due to  \citep{robins1995} works as follows.\footnote{\citet[537]{kang2007} note that `[s]ome DR estimators have been known to survey statisticians since the late 1970s.'}
The idea is to estimate the APO via
\begin{align} \label{eq:dr_target}
    p_{dr}(x,t) :=  p(x,t) + w(x,t) \frac{1}{|\JJ^x|} \sum_{i \in \JJ_t^x} \left( y_i - p(x,t) \right).
\end{align}
Here, we need to assume either that
\begin{align} \label{eq:doubly_CIA}
    \sum_{x \in \XX} \frac{|\JJ^x|}{|\JJ|} \left( \avgIIytx - \avgJJytx \right) \cdot w(x,t)
    \approx_\delta 0,
\end{align}
or that 
\begin{align} \label{eq:doubly_CIA_2}
    \sum_{x \in \XX} \frac{|\II^x|}{|\II|} \left( \avgIIytx - \avgJJytx \right)
    \approx_\delta 0.
\end{align}
This is justified if the signed differences between $x$-wise averages can be assumed to be close to zero on average and/or not strongly correlated (as a function of $x$) with $w(x,t)$, similar to (\ref{eq:exact_match_avg_signed_diff}).
This is, again, related to but weaker than the conventional unconfoundedness assumption.\footnote{Note that we typically assume that the marginal distribution over $x$ is the same for train and test, such that the difference between \ref{eq:doubly_CIA} and \ref{eq:doubly_CIA_2} in that regard is not too important.} 

Then it is sufficient to either correctly estimate the conditional means, i.e. 
\begin{equation}\label{eq:correct_means}
    \forall x \in \XX, t \in \TT: \quad
    p(x,t) = \avgIIytx,
\end{equation}
%
or to correctly estimate how often each $x$ occurs in $J_t$ as compared to $\II$ through $w$ in the sense of\footnote{One can recover the RCM version of doubly robust estimators, simply using the inverse of the empirical propensity score $w(x,t) = \frac{|\JJ^x|}{|\JJ_t^x|}$, under the (strong) assumption that $\forall x \in \XX : \frac{|\JJ^x|}{|\JJ|} = \frac{|\II^x|}{|\II|}$.} 
\begin{equation}\label{eq:correct_weights}
    \forall x \in \XX, t \in \TT: \quad
    w(x,t) = \frac{|\II^x|}{|\JJ_t^x|}\frac{|\JJ|}{|\II|}.
\end{equation}
\begin{thm}
\begin{proposition}[Predicting the APO through doubly-robust estimators]
    \ \\
    Fix any $t \in \TT$.
    Assuming $\eps$-SP (\ref{eq:SPA}), the doubly robust estimator $p_{dr}(x,t)$ provides an $(\epsilon+\delta)$-good approximation of the APO for $t$ in the sense of 
    \begin{equation}
        \left| \avgIIyt
        - \sum_{x \in \XX} \frac{|\JJ^x|}{|\JJ|} p_{dr}(x,t) \right| 
        < \epsilon + \delta,
    \end{equation}
    if either (\ref{eq:doubly_CIA}) and (\ref{eq:correct_means}), or (\ref{eq:doubly_CIA_2}) and (\ref{eq:correct_weights}) hold.
\end{proposition}
\end{thm}
\begin{proof}
    From (\ref{eq:correct_means}) and (\ref{eq:doubly_CIA}) we get 
    \begin{align}
        & \sum_{x \in \XX} \frac{|\JJ^x|}{|\JJ|} w(x,t) \frac{1}{|\JJ^x|} \sum_{i \in \JJ_t^x} \left( y_i - p(x,t) \right)  \nonumber \\
        &= \sum_{x \in \XX} \frac{|\JJ^x|}{|\JJ|} w(x,t) \left( \frac{1}{|\JJ^x|} \sum_{i \in \JJ_t^{x}} y_i - \frac{1}{|\II^x|} \sum_{i \in \II^x} \y(i,t) \right) \\
        &= \sum_{x \in \XX} \frac{|\JJ^x|}{|\JJ|} w(x,t) \left( \avgJJytx - \avgIIytx \right) \\
        &\approx_\delta 0.
    \end{align}
    Therefore, using $\eps$-SP (\ref{eq:SPA}) once again, we get
    \begin{align}
        \sum_{x \in \XX} \frac{|\JJ^x|}{|\JJ|} p_{dr}(x,t)
        &= \sum_{x \in \XX} \frac{|\JJ^x|}{|\JJ|} \left( p(x,t) +  w(x,t) \frac{1}{|\JJ^x|} \sum_{i \in \JJ_t^x} \left( y_i - p(x,t) \right) \right) \\
        &\approx_\delta \sum_{x \in \XX} \frac{|\JJ^x|}{|\JJ|} \left( p(x,t) + 0 \right) \\
        &= \frac{1}{|\JJ|} \sum_{i \in \JJ} p(x_i,t) \\
        &\approx_\eps  \frac{1}{|\II|}\sum_{i \in \II} p(\x(i),t) \label{eq:dr_mean_2-1} \\
        &=  \frac{1}{|\II|} \sum_{i \in \II} \y(i,t) = \avgIIyt. \label{eq:dr_mean_2-2}
    \end{align}
    %
    %
    %
    Alternatively, if (\ref{eq:correct_weights}) holds instead of (\ref{eq:correct_means}), we can derive
    \begin{align}
        \avgJJpt &=  \sum_{x \in \XX} \frac{|\JJ^x|}{|\JJ|} p_{dr}(x,t) \\
        &=  \sum_{x \in \XX} \frac{|\JJ^x|}{|\JJ|} \left( p(x,t) + w(x,t) \frac{1}{|\JJ^x|} \sum_{i \in \JJ_t^x} \left( y_i - p(x,t) \right) \right) \\
        &=  \sum_{x \in \XX} \frac{|\JJ^x|}{|\JJ|} \left( p(x,t) + \frac{|\JJ|}{|\JJ^x|} \frac{|\II^x|}{|\II|} \frac{1}{|\JJ_t^x|} \sum_{i \in \JJ_t^x} \left( y_i - p(x,t) \right) \right) \\
        &=  \sum_{x \in \XX} \frac{|\JJ^x|}{|\JJ|} \left( p(x,t) - \frac{|\JJ|}{|\JJ^x|} \frac{|\II^x|}{|\II|} p(x,t) + \frac{|\JJ|}{|\JJ^x|} \frac{|\II^x|}{|\II|} \frac{1}{|\JJ_t^x|} \sum_{i \in \JJ_t^{x}} y_i  \right) \\
        &=  \sum_{x \in \XX} \frac{|\JJ^x|}{|\JJ|} p(x,t) - \sum_{x \in \XX} \frac{|\II^x|}{|\II|} p(x,t) + \sum_{x \in \XX} \frac{|\II^x|}{|\II|} \frac{1}{|\JJ_t^x|} \sum_{i \in \JJ_t^{x}} y_i \\
        &\approx_\eps  \sum_{x \in \XX} \frac{|\II^x|}{|\II|} \frac{1}{|\JJ_t^x|} \sum_{i \in \JJ_t^{x}} y_i \\
        &=  \sum_{x \in \XX} \frac{|\II^x|}{|\II|} \avgJJytx \\
        &\approx_\delta  \sum_{x \in \XX} \frac{|\II^x|}{|\II|} \avgIIytx 
        = \avgIIyt.
    \end{align}
    The first approximation uses $\eps$-SP, whereas the second approximation uses (\ref{eq:doubly_CIA}).
\end{proof}

In this sense, the estimator is doubly robust.
But both (\ref{eq:correct_means}) and (\ref{eq:correct_weights}) are clearly very strong assumptions.
Double ML \citep{chernozhukov2018} assumes that we converge to this ideal eventually.
But the goal of Machine Learning is to minimise average prediction loss, not to identify true conditional distributions.
Hence, it seems that double-ML, as double robust estimators, is still limited in its applicability, and statements such as the following may apply to double-ML as well:
`There are papers suggesting that under some circumstances, estimating a shaky causal model and a shaky selection model should be doubly robust. Our results indicate that under other circumstances, the technique is doubly frail' \citep[401]{freedmanberk2008}.


\subsection{Non-compliance: Instrumental variables}

In many settings, no randomness or unconfoundedness assumption can be justified for the treatment assignment.
Sometimes, an instrumental variable is available that can be seen as random and stands in a particular relationship to the treatment assignment of interest.
A classic example from \citep{angrist1990} is the draft lottery as an instrument for examining the effect of military service on earnings.
Consider then an instrumental variable with values in $\ZZ = \{0,1\}$ and assume that we have a predictor $\py: \ZZ \times \XX \to \YY$ satisfying $\delta$-CTP (\ref{eq:CTP}) for $z$-wise (rather than $t$-wise) predictions, in the sense that
\begin{equation}\label{eq:IV_CFD}
    \forall z \in \ZZ:\quad
     \frac{1}{|\II|} \sum_{i \in \II} \y(i, z)
    \approx_\delta \frac{1}{|\II|} \sum_{i \in \II} \py(z, \x(i)).
\end{equation}
This could be justified by one of the approaches discussed so far.
%
If we have a setup where $z$ is essentially random, as in a lottery, we can simply define a predictor $\py: \ZZ \to \YY$ based on the average outcome per $\JJ_z$ as in RCTs, i.e.
\begin{equation}\label{eq:IV_bin_py_avg}
    \forall z \in \ZZ:\quad
    \py(z) := \frac{1}{|\JJ_z|} \sum_{i \in \JJ_z} y_i
    \approx_\delta \frac{1}{|\II|} \sum_{i \in \II} \y(i,z).
\end{equation}
%
Now further assume that
\begin{equation}\label{eq:exclusion}
    \forall z \in \ZZ, t \in \TT:\quad
    \forall i \in \II_{tz}: \y(i, z=z) = \y(i, t=t),
\end{equation}
where
\begin{equation}
    \II_{tz} := \{i \in \II : \s(i, z) = t\}
\end{equation}
is unknown.
This echoes the common assumption that $z$ affects $y$ only through $t$, called the `exclusion restriction' \citep{angrist1996}.
In our framework, it means that, in considering predictions for $y$, our model treats interventions on $t$ exactly as it does values of $t$ when $z$ is intervened upon (or assigned randomly in the setup).
Note that our groups $\II_{tz}$ differ from the compliance groups in the LATE estimates \citep{imbens1994} in that they a) partition only the future population $\II$ and b) do so in two pairs, sorted by which treatment $t \in \{0,1\}$ they take given assignment $z$: $\II_{00} \cup \II_{10} = \II = \II_{01} \cup \II_{11}$.

For a potentially novel result, we further assume what we shall call `dominance', namely that
\begin{equation}\label{eq:IV_monoton_01}
    \sum_{i \in \II_{01}} \y(i, t=0) 
    \leq \sum_{i \in \II_{01}} \y(i, t=1)
\end{equation} 
and
\begin{equation}\label{eq:IV_monoton_10}
    \sum_{i \in \II_{10}} \y(i, t=0) 
    \leq \sum_{i \in \II_{10}} \y(i, t=1).
\end{equation} 
The idea here is that for large enough groups, we are confident that the treatment has no negative effect on the average outcome.
This is not always sensible and needs to be justified for each case -- one needs to already have some qualitative understanding of the treatments.
Based on this, we can derive the following result:
\begin{thm}
\begin{proposition}[Lower bound on the ATE with IVs]
    \ \\
    Assume $\eps$-SP (\ref{eq:SPA}) and $\delta$-CFD (\ref{eq:IV_CFD}) for some predictor $\py$, as well as the exclusion restriction (\ref{eq:exclusion}) and dominance (\ref{eq:IV_monoton_01}), (\ref{eq:IV_monoton_10}).
    Then we can lower-bound the ATE via
    \begin{align}
        \mu(\II,t=1) - \mu(\II,t=0)\
        \geq \
        \rho(\JJ, z=1) - \rho(\JJ, z=0)
        - 2 (\eps + \delta).
    \end{align}
\end{proposition}
\end{thm}
\begin{proof}
    Using (\ref{eq:IV_monoton_01}) and (\ref{eq:IV_CFD}), we can derive
    \begin{align}
        \frac{1}{|\II|} \sum_{i \in \II} \py(\x(i),z=1)
        &\approx_\delta \frac{1}{|\II|} \sum_{i \in \II} \y(i, z=1) \\
        &=  \frac{1}{|\II|} \sum_{i \in \II_{01}} \y(i, t=0)
        + \frac{1}{|\II|} \sum_{i \in \II_{11}} \y(i, t=1) \\
        &\leq \frac{1}{|\II|} \sum_{i \in \II_{01}} \y(i, t=1)
        + \frac{1}{|\II|} \sum_{i \in \II_{11}} \y(i, t=1)\\
        &=  \frac{1}{|\II|} \sum_{i \in \II} \y(i, t=1).
    \end{align}
    For (\ref{eq:IV_monoton_10}), we analogously get    
    \begin{align}
        \frac{1}{|\II|} \sum_{i \in \II} \py(\x(i),z=0) + \delta
        &\geq \frac{1}{|\II|} \sum_{i \in \II} \y(i, t=0).
    \end{align}
    Hence, we can lower-bound the difference in APOs by 
    \begin{align}
        \frac{1}{|\II|} \sum_{i \in \II} &\y(i, t=1)
        - \frac{1}{|\II|} \sum_{i \in \II} \y(i, t=0) + 2 \delta\\
        &\geq 
        \frac{1}{|\II|} \sum_{i \in \II} \py(\x(i),1)
        - \frac{1}{|\II|} \sum_{i \in \II} \py(\x(i),0).
    \end{align}
    With the usual $\eps$-SP assumption, we thus get
    \begin{align}
        \frac{1}{|\II|} \sum_{i \in \II}& \y(i, t=1) - \frac{1}{|\II|} \sum_{i \in \II} \y(i, t=0) + 2 \eps + 2 \delta \\
        &\geq 
        \frac{1}{|\JJ|} \sum_{i \in \JJ} \py(x_i,z=1) - \frac{1}{|\JJ|} \sum_{i \in \JJ} \py(x_i,z=0).
    \end{align}
\end{proof}
In the special case where the instrument is assigned randomly and we can assume (\ref{eq:IV_bin_py_avg}), we get the following.
\begin{thm}
\begin{corollary}[Lower bound on the ATE with randomised IVs]
    \ \\
    Assume $\eps$-SP (\ref{eq:SPA}) and (\ref{eq:IV_bin_py_avg}), as well as the exclusion restriction (\ref{eq:exclusion}) and dominance (\ref{eq:IV_monoton_01}), (\ref{eq:IV_monoton_10}).
    Then we can lower-bound the ATE via
    \begin{align}
        \mu(\II,t=1) - \mu(\II,t=0)\
        \geq\
        \mu(\JJ,t=1) - \mu(\JJ,t=0)\
        - 2 (\eps + \delta).
    \end{align}
\end{corollary}
\end{thm}
\begin{proof}
    We can use the above Proposition and simply insert $\py$ as in (\ref{eq:IV_bin_py_avg}).
    Then we get
    \begin{align}
        \frac{1}{|\II|} \sum_{i \in \II}& \y(i, t=1) - \frac{1}{|\II|} \sum_{i \in \II} \y(i, t=0)  + 2 \eps + 2 \delta\\
        &\geq 
        \py(1) - \py(0)
        = \frac{1}{|\JJ_1|} \sum_{i \in \JJ_1} y_i - \frac{1}{|\JJ_0|} \sum_{i \in \JJ_0} y_i.
    \end{align}
\end{proof}

Two differences to the LATE methodology are worth noting:
First, LATE estimates supposedly identify the average treatment effect on the group of `compliers' \citep{angrist1996}, which for us is the group $\II_{00} \cap \II_{11}$.
For this to make sense, we would need to assume that these groups are well-defined even if the respective treatment is not assigned.
This makes these statements metaphysically strong and unverifiable in principle.
This is related to what \citet{dawid2000} calls `fatalism', namely the assumption `that the various potential responses $Y_{ti}$, when treatment $t$ is applied to unit $i$, as predetermined attributes of unit $i$, waiting only to be uncovered by suitable experimentation' (p. 412, notation adapted).
Considering the LATE estimates that are usually ascribed to compliers, he notes that `it is only under the unrealistic assumption of fatalism that this group has any meaningful identity, and thus only in this case could such inferences even begin to have any useful content' (p. 413).
We agree that these groups are not well-defined, as counterfactuals are not.
Ascribing specific LATEs to supposedly fixed subgroups defined by counterfactuals thus relies on strong metaphysical assumptions; and typically, we are interested not in such subgroups (even if they were well-defined), but in the whole population \citep{deaton2009, heckman2010} -- which is a practical reason to focus on lower bounds in the ATE of the whole population.
For our approach, the outcomes also need not be well-defined before the treatment is assigned.
The estimation of the ATE could not be directly tested, but it could be tested by randomly assigning two treatments to a future group and observing the population averages.
Not even this is possible with LATE, as the group itself cannot be determined by observation.

Second, our assumptions (\ref{eq:IV_monoton_01}) and (\ref{eq:IV_monoton_10}) are very different to the typical `monotonicity' assumption \citep{imbens1994}, according to which there is nobody who would get $\s(i, z=1) = 0$ but $\s(i, z=0) = 1$, i.e. for whom higher $z$ leads to lower $t$.
Even if LATEs would make sense, it would arguably be very strong to assume that \textit{no} such person exists: it would not be enough that it increases the chance of treatment for everyone.
In contrast, (\ref{eq:IV_monoton_01}) and (\ref{eq:IV_monoton_10}) say that higher $t$ leads to higher $y$ \textit{on average}.
Hence are not only weaker but also less metaphysical, as they do not involve counterfactuals.
Unfortunately, they are still untestable -- which arguably makes IV approaches somewhat less reliable than other approaches -- still we show how they can be used for estimating quantities on the population level.\footnote{It has already been previously observed by \citet{robins1989} and \citet{manski1990} that -- without assuming dominance -- one can bound the APO from above and below if there is an upper and a lower bound on possible outcomes.}


\subsection{Predicting the past: Difference-in-differences}
\label{ss:d-in-d}

While the approaches so far have a clear forward-looking component, difference-in-differences, as well as synthetic control methods, are by design more backwards-looking.
Ultimately, however, all such methods are meant to inform policy making.
We already discussed in Section~\ref{ss:distribution} that the distribution framing can entice one to be overly optimistic about the generalisability of one's results.
As pointed out by \citet{deaton2018} (in the context of RCTs), a focus on internal validity `is sometimes incorrectly taken to imply that results of an internally valid trial will automatically, or often, apply ``as is'’ elsewhere, or that this should be the default assumption failing arguments to the contrary' (p.10).
In the following, we demonstrate how our framework makes sense of diff-in-diff approaches and, in doing so, directly involves the question of generalisation or induction (the case for synthetic control methods is similar).

We consider the following setting:
Assume that there are three groups $\GG = \{A,B,C\}$; we have data for group $A$ and $B$ and want to inform treatment decisions about group $C$.
The data concerns two steps $\Scal = \{0,1\}$; our data includes treatment $t=1$ at step $s=1$ for group $A$ and treatment $t=0$ at step $s=1$ for group $B$.
We will consider groups
\begin{equation*}
    \JJ_{a}^0,\ \JJ_{a1}^1,\ \JJ_{b}^0,\ \JJ_{b0}^1,\ \JJ_{c}^0,\ \II_{c}
\end{equation*}
(in the form $\JJ_{gt}^s$) with observed mean outcomes 
\begin{equation*}
    \mu(\JJ_{a}^0),\ \mu(\JJ_{a}^1, t=1),\ \mu(\JJ_{b}^0),\ \mu(\JJ_{b}^1, t=0),\ \mu(\JJ_{c}^0).
\end{equation*}
%
At step $s=0$, we do not need treatment indicators so we denote the people in the three groups by $\JJ_{a}^0, \JJ_{b}^0$ and $\JJ_{c}^0$. Lastly, $\II_{c}$ denotes the people of group $C$ at step $s=1$, for which we intend to make an informed treatment assignment.

We then assume that the group $C$ is comparable to $A$ and to $B$ in terms of the difference of averages between time steps with the same treatment, i.e.
\begin{equation}
    \mu(\II_{c}, t=1)
    - \mu(\JJ_{c}^0)
    \ \approx_\eps \ \mu(\JJ_{a}^1, t=1)
    - \mu(\JJ_{a}^0)
\end{equation}
and
\begin{equation}
    \mu(\II_{c}, t=1)
    - \mu(\JJ_{c}^0)
    \ \approx_\delta \ \mu(\JJ_{b}^1, t=0)
    -\mu(\JJ_{b}^0).
\end{equation}
We can, thus, predict the APOs as 
\begin{equation}
    \mu(\II_{c}, t=1)
    \ \approx_\eps \ \mu(\JJ_{a}^1, t=1)
    - \mu(\JJ_{a}^0)
    + \mu(\JJ_{c}^0).
\end{equation}
and
\begin{equation}
    \mu(\II_{c}, t=0)
    \ \approx_\delta \ \mu(\JJ_{b}^1, t=0)
    - \mu(\JJ_{b}^0)
    + \mu(\JJ_{c}^0).
\end{equation}

In comparison to standard accounts of diff-in-diff, we have directly included the group $C$ that we want to make predictions about, rather than trying to infer supposed causal relationships in $A$ or $B$.
If we want to inform policymaking through our modelling, then we need to think about a different group $C$.
There may, however, also be reasons to make (unverifiable) predictions about what would have happened in $A$ under $t=0$ or in $B$ under $t=1$.
For this, we can use the model by inserting $A$ or $B$ for $C$.
Note that we do not, strictly speaking, \textit{learn} anything about what \textit{would} have happened:
We merely make a (potentially well-founded) prediction -- a prediction that is unverifiable in principle.
Still, it may provide an argument for or against other models:
In the well-known case of \citep{card1994} concerning the minimum wage in New Jersey and Pennsylvania, the interesting insight is that their model contradicts the general economics model that predicts lower employment for higher wages.
While their analysis does not constitute \textit{empirical} evidence against the general model, it can -- if considered well-justified -- still provide a strong reason to doubt it.


\section{Connection to non-probability sampling}
\label{app:non_prob}

Considering a treatment group $\JJ_t$ as a subsample of the observed sample $\JJ$ connects it to problems of survey sampling or missing data.
Causal inference has been considered a missing data problem from the start by Rubin, see e.g. \citep{rubin1976, little2020}.
The connections are particularly clear in finite population settings.
Inverse probability weighting was developed for finite population survey sampling \citep{horvitz1952} -- for cases when the propensity score is known by design.\footnote{The strength of the assumption that there is a well-defined propensity score -- part of the `missing at random' assumption in \citep{rubin1976} -- seems to be hardly discussed anymore, as it directly follows from the way the problem is set up.\label{fn:missing-at-random}}
\citet{abadie2020} compute standard errors for estimating causal population means in terms of the uncertainty introduced by the random assignment of treatment in the observed data.
They assume that the counterfactual outcomes are well-defined, but it should be possible to apply similar reasoning to the account pursued here.
While survey sampling is not mentioned in the cited work, the authors do draw that connection in an earlier version \citep{abadie2014}.
Conversely, the non-probability survey sampling literature in particular \citep{elliott2017, wu2022} draws heavily on early work by Rubin and others.
This literature is concerned with inference from non-representative samples from finite populations -- more precisely, from `samples without an identified design probability construct' \citep[341]{meng2022}.
Nowadays these problems are mostly discussed independently, with some exceptions.
\citet{mercer2017} explicitly construes non-probability survey sampling as a causal inference problem.

We go the opposite route and suggest seeing causal inference as an instance of predictions under non-probability sampling.
\citet{kang2007} note that `the methods described in this article [for estimating a population mean from incomplete data] can be used to estimate an average causal effect by applying the method separately to each potential outcome' (p. 525).
In contrast to this and the above-mentioned work by Abadie and colleagues, however, we see the aim in making treatment-wise predictions rather than inferences about counterfactual outcomes or individual effects.
That means that there is no so-called `fundamental problem of causal inference' \citep{holland1986}: knowing two mutually exclusive potential outcomes for some input would not help to make predictions.
The basic idea is to build predictors which take treatment as just another attribute, but one whose distribution may change dramatically in the future.
This means that we can see causal inference methods as treatment-wise predictors of potential outcomes (Figure~\ref{fig:ours}).


\section{Linear regression}
\label{app:lin_reg}

So far, we have focussed on causal inference for binary treatments, as particularly relevant for programme evaluation, but it can also be applied to settings where treatment can take multiple values.
The most popular model for such settings is linear regression.
In their landmark textbook, \citet[52]{angrist2009} vaguely note that `[a] regression is causal when the [true distributional model] it approximates is causal'.
Regression for causal inference is slightly more controversial than for binary treatments, as it needs stronger assumptions that are nevertheless less visible.
For example, Michael Freedman\footnote{See also his critique of causal regression in \citep{freedman2006, freedman2008}.} notes that
\begin{quote}
    `Lurking behind the typical regression model will be found a host of such assumptions; without them, legitimate inferences cannot be drawn from the model. There are statistical procedures for testing some of these assumptions. However, the tests often lack the power to detect substantial failures.' \citep[33]{freedman1995}
\end{quote}
These caveats become more evident when showing how linear regression fits into our framework.


\subsection{Identifying correct models}

We start by demonstrating that, assuming there is a correct linear model, regression can identify this model.
In our framework, this means that we assume there is a linear model 
\begin{equation}
    p(x,s) = a^* x + \beta^* s + c^*
\end{equation} 
that can describe the future average well for every $x \in \XX$ (analogous to assuming a linear CEF) in the sense that
\begin{equation}\label{eq:LR_true_model}
    \forall x \in \XX, t \in \TT: \quad
    \frac{1}{|\II^x|} \sum_{i \in \II^x} \y(i,s) 
    \approx_\eps \frac{1}{|\II^x|} \sum_{i \in \II^x} a^* x + \beta^* s + c^*.
\end{equation} 
One sufficient assumption is that the $\JJ_t^x$ are comparable with the $\II^x$ in the sense that the residuals are not biased, i.e.
\begin{equation}\label{eq:regression_causal}
    \forall x \in \XX, t \in \TT: \quad
    \frac{1}{|\JJ_t^x|} \sum_{i \in \JJ_t^x} y_i - p(x,t)
    \approx_\delta \frac{1}{|\II|} \sum_{i \in \II^x} \y(i, t) - p(\x(i),t)
\end{equation}
Then clearly 
\begin{equation}\label{eq:LR_identification}
    \forall x \in \XX, t \in \TT: \quad
    \frac{1}{|\JJ_t^x|} \sum_{i \in \JJ_t^x} y_i
    \approx_{\eps+\delta} \frac{1}{|\JJ_t^x|} \sum_{i \in \JJ_t^x} a^* x + \beta^* t + c^*.
\end{equation}
%
Alternatively, if the $x_i$ are uncorrelated with the $t_i$ in our data, then to show (\ref{eq:LR_identification}), it is also sufficient -- instead of (\ref{eq:regression_causal}) -- to assume only $t$-wise comparability, i.e.
\begin{equation}
    \forall t \in \TT: \quad
    \frac{1}{|\II|} \sum_{i \in \II} \y(i, t) - \frac{1}{|\II|} \sum_{i \in \II} p(\x(i),t) 
    \approx_\delta \frac{1}{|\JJ_t|} \sum_{i \in \JJ_t} y_i - \frac{1}{|\JJ_t|} \sum_{i \in \JJ_t} p(x_i,t).
\end{equation}
In this case, note that (\ref{eq:LR_true_model}) implies
\begin{align}
    \forall s,t \in \TT:
    &\frac{1}{|\II|} \sum_{i \in \II} \y(i, s)
    - \frac{1}{|\II|} \sum_{i \in \II} \y(i, t) \\
    &\approx_{2\eps} a^* \x(i) + \beta^* s + c^* - a^* \x(i) + \beta^*t + c^* \\
    &= \beta^* \cdot (s - t)
\end{align}
and thus 
\begin{align}\label{eq:OVB_in_action}
    \forall s,t \in \TT:
    \frac{1}{|\JJ_s|} \sum_{i \in \JJ_s} y_i
    - \frac{1}{|\JJ_t|} \sum_{i \in \JJ_t} y_i
    &\approx_{2 \delta + 2 \eps} \beta^* \cdot (s - t),
\end{align}
%

In the case of perfect fits, i.e. $\eps = \delta = 0$ fitting a linear model with MSE loss would identify the true model, as MSE elicits the mean.
Furthermore, in the case of the $x_i$ being uncorrelated with the $t_i$, the OVB formula tells us that the $\beta$ estimator in the long regression is equal to the estimator in the regression $y = \beta \cdot t$, such that either of them would identify the true parameter.
The fact that the assumptions in this subsection are very strong can be connected back to the cited critique by David Freedman -- despite the identification and validity results for linear regression that can be derived in expectation, or in the limit of infinite data.\footnote{It is also worth noting that linear models are often chosen less because there are good reasons to believe in linear models and more because they are nice to work with.}


\subsection{Instrumental variables}
\label{ss:IV_reg}

Here, we assume the assignment of the instrument $z$ was `random' in the sense that
\begin{align}\label{eq:IV_reg_random_z}
    \forall z \in \ZZ: \quad
    \frac{1}{|\II|} \sum_{i \in \II} \y(i, z)
    \approx_\delta \frac{1}{|\JJ_z|} \sum_{i \in \JJ_z} y_i.
\end{align}
%
Now further assume (as usual for IV models) that $z$ affects $y$ only through $t$ (`exclusion restriction') in the sense that
\begin{equation}\label{eq:IV_reg_excl}
    \forall i \in \II, z \in \ZZ: \quad
    \y(i, z=z) = \y(i, \s(i, z)).
\end{equation}
%
Assume further that the average outcome $y$ of an intervention on $t$ depends on $t$ only via its average, i.e. that for any treatment assignment $\dec, \dec' : \II \to \TT$ if it holds that
\begin{equation}
    \frac{1}{|\II|} \sum_{i \in \II} \dec(i)
    \approx_\gamma \frac{1}{|\II|} \sum_{i \in \II} \dec'(i)
\end{equation}
implies
\begin{align}\label{eq:IV_reg_dec}
    \frac{1}{|\II|} \sum_{i \in \II} \y(i, \dec(i))
    &\approx_\eps \frac{1}{|\II|} \sum_{i \in \II} \y(i, \dec'(i)).
\end{align}
This means in particular that outcomes are on average linear in the treatment -- which is also part of the conventional assumption that there is a true causal linear model $y = a^* x + \beta^* t + c$.

Then for any treatment rule $\dec_z : \II \to \TT$ that roughly leads to the same average treatment as assigning $z$ would do, i.e.
\begin{equation}\label{eq:dec_z}
    \frac{1}{|\II|} \sum_{i \in \II} \dec_z(i) 
    \approx_\gamma \frac{1}{|\II|} \sum_{i \in \II} \s(i,z),
\end{equation}
we know via (\ref{eq:IV_reg_dec}), (\ref{eq:IV_reg_excl}), and (\ref{eq:IV_reg_random_z}) that we can predict the average outcome of a treatment rule $\dec_z$ as 
\begin{equation}
    \frac{1}{|\II|} \sum_{i \in \II} \y(i, \dec_z(i))
    \approx_\eps \frac{1}{|\II|} \sum_{i \in \II} \y(i, \s(i,z))
    = \frac{1}{|\II|} \sum_{i \in \II} \y(i, z)
    \approx_\delta \frac{1}{|\JJ_z|} \sum_{i \in \JJ_z} y_i.
\end{equation}
To use this, we need to know which instrument would have led to a similar average treatment.
For this, we can investigate the data under the assumption that
\begin{equation}\label{eq:IV_reg_ps}
    \forall z \in \ZZ: \quad
    \frac{1}{|\II|} \sum_{i \in \II} \s(i,z)
    \approx \frac{1}{|\JJ_z|} \sum_{i \in \JJ_z} t_i,
\end{equation}
justified by the random assignment of $z$.

This leads to a procedure that is somewhat reminiscent of the two steps in 2SLS:
To estimate the APO of a treatment $t$, we first analyse the observed relationship between instrument and treatment to find a $z \in \ZZ$ with (\ref{eq:dec_z}).
Then we use the observed relationship between instrument and outcome to predict the APO of the treatment.
The main difference to 2SLS is that we do not try to identify parameters of an assumed linear model here, which makes it more general.
Also note that our approach can be straightforwardly generalised to predict the APO of a policy $\pol : \XX \to \TT$ instead of a constant treatment $t$ in cases where we have access to covariates $\XX$.


\section{Predicting quantiles}
\label{app:quantiles}

We here elaborate on the prediction of quantiles of the outcome distribution, 
\begin{equation}
    \qtp := \min \left\{ y \in \YY :  F_{\II,t}(y) \geq p \right\},
\end{equation}
as briefly discussed in Section~\ref{ss:beyond_average}.
For RCTs where it is justified to assume no bias between sample and target population, we can simply take the observed quantile in group $\JJ_t$ as an estimator.
This can be justified when $\JJ_t$ and $\II$ can be considered as making up the same population, which means that the proportion $p$ of values below a certain threshold should be similar in both groups.
We denoting the $p$-quantile in $\JJ_t$ as
\begin{equation}
    \qtpJJ := \min \left\{ y \in \YY :  F_{\JJ_t}(y) \geq p \right\}.
\end{equation}
We denote
\begin{equation}
    \gamma := |F_{\JJ_t}(y) - p|
\end{equation}
which is measurable (and indeed we can tweak $p$ for $\gamma$ to be zero).
Then we formulate the dummy outcomes 
\begin{equation*}
    \yitil := \mathbb{1}[y_i \leq \qtpJJ] 
    \quad \text{and} \quad
    \ytil(i,t) := \mathbb{1}[\y(i,t) \leq \qtpJJ].
\end{equation*}
Then from the unbiasedness assumption
\begin{equation}
    F_{\II,t}(\qtpJJ) = \frac{1}{|\II|} \sum_{i \in \II} \ytil(i,t) \approx_\delta \frac{1}{|\JJ_t|} \sum_{i \in \JJ_t} \yitil = F_{\JJ_t}(\qtpJJ),
\end{equation}
analogous to (\ref{eq:RCT-APO}), we get that
\begin{equation}
    F_{\II,t}(\qtpJJ) \approx_\delta F_{\JJ_t}(\qtpJJ) \approx_\gamma p.
\end{equation}
Then the bounded variability assumption (\ref{eq:local_lipschitz}) lets us bound
\begin{equation}
    | \qtpJJ - \qtp | < \alpha(\delta+\gamma).
\end{equation}

For non-RCT samples, we can use an approach similar to matching or inverse probability weighting.
It has been previously observed that one can use a version of inverse probability weighting to estimate the cumulative outcome distribution.
For both fine-and coarse grained (i.e. stratified) approaches, it is common to use parametric propensity score models, as in \citet{zhang2012}.
In line with out previous discussion of exact matching (and our discussion of coarse-grained exact matching in the Appendix), we discuss the use of empirical propensity scores here.
The idea is again to weight instances for treatment $t$ (that is, in group $\JJ_t$) based on their occurrence of their covariates in the entire sample $\JJ$.
The estimator $\qtpHat$ is then defined as
\begin{equation}
    \qtpHat := 
    \min \left\{ y \in \YY :
    \frac{1}{|\JJ|} \sum_{i \in \JJ_t} \frac{\mathbb{1}[y_i \geq y]}{\prop_t(x_i)} 
    \geq p \right\},
\end{equation}
where $\prop_t(x) := \frac{|\JJ_t^x|}{|\JJ^x|}$ is again the observed propensity score for treatment $t$.
%
Now we define dummy outcomes w.r.t. $\qtpHat$, as
\begin{equation*}
    \yitil := \mathbb{1}[y_i \leq \qtpHat] 
    \quad \text{and} \quad
    \ytil(i,t) := \mathbb{1}[\y(i,t) \leq \qtpHat].
\end{equation*}
Similar to above, we define 
\begin{equation}
    \gamma := \left|\frac{1}{|\JJ|}\sum_{i \in \JJ_t} \frac{\yitil}{\prop_t(x_i)} - p\right|,
\end{equation}
which is usually small.
Then, similar to Section~\ref{ss:matching}, we assume that the \textbf{average (signed) difference} between the $x$-wise proportions of outcomes below $\qtpHat$,
\begin{equation*}
     r_x^q(\II,t) := \frac{1}{|\II^x|} \sum_{i \in \II^x} \ytil(i,t)
    \quad \text{and} \quad 
     r_x^q(\JJ,t) := \frac{1}{|\JJ_t^x|} \sum_{i \in \JJ_t^x} \yitil,
\end{equation*}
is not strongly biased above or below zero, that is,
\begin{equation}\label{eq:quant_exact_match_avg_signed_diff}
    \left|  \sum_{x \in \XX} \frac{|\II^x|}{|\II|} \left( r_x^q(\II,t) - r_x^q(\JJ,t) \right) \right|
    < \delta.
\end{equation}
%
Analogous to $\eps$-SAP, we assume that the distribution on $\XX$ is similar on $\JJ$ compared to $\II$ in the sense that 
\begin{equation}
     \sum_{x \in \XX} \frac{|\II^x|}{|\II|} r_x^q(\JJ,t) \approx_\eps \sum_{x \in \XX} \frac{|\JJ^x|}{|\JJ|} r_x^q(\JJ,t).
\end{equation}
Now since
\begin{equation}
     F_{\II,t}(\qtpHat)
     = \frac{1}{|\II|} \sum_{i \in \II} \ytil(i,t) 
     = \sum_{x \in \XX} \frac{|\II^x|}{|\II|} r_x^q(\II,t),
\end{equation}
and
\begin{equation}
     \sum_{x \in \XX} \frac{|\JJ^x|}{|\JJ|} r_x^q(\JJ,t) = \frac{1}{|\JJ|}\sum_{i \in \JJ_t} \frac{\yitil}{\prop_t(x_i)} \approx_\gamma p,
\end{equation}
by definition of $\gamma$, this gives us
\begin{equation}
    F_{\II,t}(\qtpHat) 
    \approx_\delta \sum_{x \in \XX} \frac{|\II^x|}{|\II|} r_x^q(\JJ,t) 
    \approx_\eps \sum_{x \in \XX} \frac{|\JJ^x|}{|\JJ|} r_x^q(\JJ,t)
    \approx_\gamma p.
\end{equation}
As above, via (\ref{eq:local_lipschitz}) we can then bound
\begin{equation}
    | \qtpHat - \qtp | < \alpha(\eps+\delta+\gamma).
\end{equation}



\end{document}